\documentclass[iop]{emulateapj}
\usepackage{graphicx}
\usepackage{amsmath}
\newcommand\m{\mathrm}
\newcommand\beq{\begin{equation}}
\newcommand\eeq{\end{equation}}
\newcommand\fobs{$\langle f_\m{obs} \rangle$}
\allowdisplaybreaks

\begin{document}

\title{The Direct Detectability of Giant Exoplanets in the Optical}
\author{Johnny P. Greco$^\star$ and Adam Burrows$^\dagger$} 
\email{$^\star$jgreco@astro.princeto.edu}
\email{$^\dagger$burrows@astro.princeton.edu}
\affil{Department of Astrophysical Sciences, Princeton University, Princeton, NJ 08544}
\shorttitle{The Direct Detectability of Giant Exoplanets in the Optical}
\shortauthors{Greco \& Burrows}
\begin{abstract}  
Motivated by the possibility that a coronagraph will be put on WFIRST/AFTA, we explore the direct detectability of extrasolar giant planets (EGPs) in the optical. We quantify a planet's detectability by the fraction of its orbit for which it is in an observable configuration ($f_\m{obs}$). Using a suite of Monte Carlo experiments, we study the dependence of $f_\m{obs}$ upon the inner working angle (IWA) and minimum achievable contrast ($C_\m{min}$) of the direct-imaging observatory; the planet's phase function, geometric albedo, single-scattering albedo, radius, and distance from Earth; and the semi-major axis distribution of EGPs. We calculate phase functions for a given geometric or single-scattering albedo, assuming various scattering mechanisms. We find that the Lambertian phase function can predict significantly larger $f_\m{obs}$'s with respect to the more realistic Rayleigh phase function. For observations made with WFIRST/AFTA's baseline capabilities ($C_\m{min}\sim10^{-9}$, $\m{IWA}\sim0.2''$), Jupiter-like planets orbiting stars within 10, 30, and 50 parsecs of Earth have volume-averaged observability fractions of ${\sim}$12\%,\ 3\%, and 0.5\%, respectively. At 10 parsecs, such observations yield $f_\m{obs}>1\%$ for low- to modest-eccentricity planets with semi-major axes in the range ${\sim}2 - 10$~AU. If $C_\m{min}=10^{-10}$, this range extends to ${\sim}35$~AU. We find that, in all but the most optimistic configurations, the probability for detection in a blind search is low (${<}\,5\%$). However, with orbital parameter constraints from long-term radial-velocity campaigns and {\it Gaia} astrometry, the tools we develop in this work can be used to determine both the most promising systems to target and when to observe them.
\end{abstract}


\section{Introduction}
In the burgeoning field of exoplanetary science, the scientific return from direct imaging has so far been meager when compared to the very successful radial-velocity and transit methods. This delay in progress has been primarily due to the technological challenge of performing high-contrast observations at small angular separations \citep{Oppenheimer2009, Truab2010}. For a Jupiter twin, the plant/star flux ratio is expected to be ${\sim}10^{-7}$ in the mid-infrared and ${\sim}10^{-9}$ in the optical, and if the planet is in close proximity to Earth ($<\!30$~pc), these contrasts must be achieved at angular separations of a few tenths of arcseconds. Making the problem harder still, the direct detectability of a given planet will vary with time, depending upon its Keplerian orbital elements and atmospheric properties---which will depend upon the planet's age, mass, orbital distance, and parent star \citep{Sudarsky2005, Kane2010, Kane2011, Madhu2012}.

Despite these challenges, recent progress has been made via high-contrast observations in the near-infrared of nearby, young extrasolar giant planets (EGPs)/brown dwarfs at very wide orbital separations (${\sim}10-200$~AU). Such substellar companions are self-luminous and provide contrasts and angular separations that are far more favorable than the case of a nearby Jupiter twin. Indeed, there have been a number of notable discoveries and partial characterizations, including 2MASS~1207b \citep{Chauvin2004}, AB~Pic~b \citep{Chauvin2005}, GQ~Lup~b \citep{GQLupb}, $\beta$-Pictoris~b \citep{Lagrange2009}, HR~8799bcde \citep{Marois2008, Marois2010, Barman2011}, and GJ~504b \citep{Kuzuhara2013}. In the coming years, a plethora of high-contrast instruments such as the ground-based GPI \citep{Macintosh2008}, SPHERE \citep{Beuzit2008}, LBTI \citep{Skrutskie2010}, ScExAO/Charis/HiCIAO \citep{Suzuki2010} and the space-based NIRCam and MIRI on JWST \citep{Deming2009} will almost certainly produce a rich sample of directly imaged and characterized young EGPs  \citep{Beichman2010}.  

All of the above high-contrast imaging efforts are focused on measurements in the infrared. However, optical measurements of albedo spectra, phase curves, and polarization---which have been used to infer chemical compositions and cloud properties of several planets and moons in the solar system \citep{Hansen1974, Karkoschka1994, Satoh2000, Irwin2002, Karkoschka2011}---have enormous potential to constrain exoplanet atmospheric compositions, Keplerian elements, and aerosol and cloud properties \citep{Marley1999, Sudarsky2000, Sudarsky2003, Sudarsky2005, Burrows2004, Seager2000, Stam2004,  Dyudina2005, Buenzli2009, Cahoy2010, Madhu2012, Burrows2014, Marley2014}. Additionally, optical reflected-light observations are sensitive to cool, mature EGPs at wide orbital distances, a population of planets that has been inaccessible to infrared measurements. Direct imaging and characterization of cool, mature EGPs will require a space-based high-contrast instrument such as the anticipated optical imaging and spectrophotometric coronagraph that is being designed for the Wide-Field Infrared Survey Telescope (WFIRST)/Astrophysics Focused Telescope Assets (AFTA) mission. If realized, this instrument will achieve planet/star contrast ratios better than ${\sim}10^{-9}$ and an inner working angle near or better than ${\sim}0.2''$ \citep{Spergel2015}, well within the necessary capability to detect a Jupiter twin located ${\sim}20$~pc form Earth\footnote{For a summary of the diagnostic potential for cold exoplanet characterization using the anticipated WFIRST/AFTA coronagraph, see \citet{Burrows2014}, \citet{Marley2014}, and \citet{Hu2014}.}. Therefore, studies of the observational signatures and direct detectability of reflected light from EGPs are particularly timely and will aid in the design and optimization of WFIRST/AFTA or a similar space-based effort. 

In this work, we present an exploration of the direct detectability of cool EGPs at visible wavelengths (${\sim}0.4-1.0~\m{\mu m}$). We quantify a planet's direct detectability by the fraction of its orbit for which it is in an observable configuration (its observability fraction $f_\m{obs}$). Using a suite of Monte Carlo (MC) experiments, we explore the dependence of $f_\m{obs}$ upon various input parameters such as the inner working angle (IWA), minimum achievable contrast ($C_\m{min}$), planetary albedo and radius, distance from Earth, and the semi-major axis distribution of EGPs. We assume cloud-free, homogeneous atmospheres and calculate phase functions for a given geometric or single-scattering albedo, assuming different scattering mechanisms. Although partly motivated by the possibility of an optical coronagraph on WFIRST/AFTA, our study is relevant to any effort to directly detect exoplanets in the optical.  

In \textsection\ref{sec:detect} we describe how we model reflected-light observations of EGPs. We tabulate and provide analytic fits to the geometric albedo as a function of single-scattering albedo for scalar and vector Rayleigh, Lambert, and isotropic scattering. In addition, we present representative albedo spectra for homogeneous atmospheres of cool EGPs. This section is supported by an appendix, which reviews the equations that connect the Keplerian orbital elements to the observed phase angle and time. In \textsection\ref{sec:fr_KE} we present illustrative calculations of the variation of planet/star flux ratios and angular separations with orbital parameters, which ultimately determines an exoplanet's direct detectability. In \textsection\ref{sec:MC} we describe our MC experiments and discuss the assumptions we make about the orbital parameter distributions of EGPs. 

Our results are given in \textsection\ref{sec:results}. In \textsection\ref{sec:mean_fobs} we present a parametric study of the mean observability fraction, \fobs, for many MC experiments, which are run under different sets of assumptions. In \textsection\ref{sec:probs} we highlight the fact that \fobs\ for a population of planets is proportional to the probability that a blind search for such planets will yield a detection. Equivalently, \fobs\ can be interpreted as the probability for detection, assuming all stars have one such planet. In \textsection\ref{sec:thermal} we briefly consider the optical signature of thermal emission from young EGPs. The paper is brought to close in \textsection\ref{sec:summary} with a summary of our results.

Throughout this paper, we define a ``Jupiter-like'' planet to be an EGP with the radius of Jupiter, a geometric albedo of 0.5, and a Rayleigh scattering atmosphere. Although these atmospheric properties are not consistent with Jupiter at all wavelengths, near ${\sim}0.565~\m{\mu m}$, which is at the center of one of the WFIRST/AFTA bandpasses, they provide a decent approximation to observations \citep{Karkoschka1994}. 

\section{Modeling Reflected-Light Observations of Giant Exoplanets}\label{sec:detect}

To directly detect an extrasolar planet, its dim light must be separated from the overwhelming glare of its bright parent star. For a given high-contrast imaging observation, this requires that two conditions be {\it simultaneously} satisfied: 1) the planet/star contrast ratio ($F_p/F_\star$) must be above the minimum achievable contrast ($C_\m{min}$), and 2) the planet/star angular separation ($\delta\phi$) must be both larger than the inner working angle (IWA) and smaller than the outer working angle (OWA). As a further complication, $C_\m{min}$ will typically be a function of $\delta \phi$, and both of these parameters will depend upon the wavelength of the observation. Thus, the direct detectability of an exoplanet ultimately depends upon its location in $(F_p/F_\star)-\delta\phi$ parameter space, which is a complicated function of time, orbital and atmospheric parameters, and the high-contrast imaging technology used to make the observation. In this section, we describe the methods we employ to predict the direct detectability of EGPs as they move through $(F_p/F_\star)-\delta\phi$ parameter space.

\subsection{Planet/Star Contrast Ratios: Albedos and Phase Functions}\label{sec:albedos}

At optical wavelengths (${\sim}0.4-1.0~\mu m$), the flux from a giant exoplanet is dominated by reflected starlight, provided the planet is sufficiently old and of low enough mass that its thermal emission is negligible (see \textsection\ref{sec:thermal} for a discussion of thermal emission from young EGPs). Observationally, the wavelength- and phase-dependence of this reflection is encompassed by the geometric albedo spectrum, $A_g(\lambda)$, and the classical phase function, $\Phi(\lambda, \alpha)$. Here, $\lambda$ is the photon wavelength, and $\alpha$ is the exoplanet-centric angle between the star and the observer. In terms of these parameters, the planet/star flux ratio is given by
\beq \label{eqn:fr}
\frac{F_p}{F_\star} = A_g(\lambda) \left(\frac{R_p}{d}\right)^2\Phi(\lambda, \alpha), 
\eeq
where $R_p$ is the planet radius and $d$ is the orbital distance (see appendix, Equation~(\ref{eqn:distance})). $\Phi(\alpha, \lambda)$ is normalized to unity at full phase ($\alpha=0^\circ$), which defines the geometric albedo. The planet's atmospheric composition, temperature/pressure structure, cloud/haze profiles, and scattering properties govern $A_g(\lambda)$ and $\Phi(\lambda, \alpha)$. Hence, observations of optical planet/star flux ratios, as a function of wavelength and phase angle, have great potential to constrain the theory of exoplanet atmospheres \citep[e.g.,][]{Marley1999, Sudarsky2000, Burrows2004, Cahoy2010}. Furthermore, the shape and magnitude of an exoplanet's phase-folded light curve are highly sensitive to the Keplerian elements of its orbit \citep{Sudarsky2005, Dyudina2005, Kane2010, Kane2011, Madhu2012}, providing an independent method for measuring orbital parameters that complements astrometric and radial-velocity methods.

In this work, we are interested in the general character, as opposed to detailed quantitative predictions, of the direct detectability of EGPs. We, therefore, generally assume planets with constant $A_g$ (the exception is in \textsection\ref{sec:Agdist}, where we vary $A_g$ with orbital distance using two simple models) and use the formalism presented in \citet{Madhu2012} to calculate the associated single-scattering albedo ($\omega$) and phase function for an assumed scattering mechanism. In Table~\ref{deluxetable:Agtable}, we have tabulated $A_g$ as a function of $\omega$ for scalar and vector Rayleigh scattering, Lambert reflection, and isotropic scattering. Unless stated otherwise, we will hereafter refer to vector Rayleigh scattering as Rayleigh scattering. 
 
\begin{deluxetable}{rcccc}
\centering
\tablecaption{Geometric albedos as a function of single-scattering albedo}
\tablecolumns{5}
\tabletypesize{\footnotesize}
\tablewidth{8.5cm}
\tablehead{
\multicolumn{1}{c}{$\omega$}  &  \multicolumn{4}{c}{$A_g$} \vspace{0.07cm} \\  \hline \vspace{-0.2cm}\\
\colhead{}& 
\colhead{} &
\colhead{} &
\colhead{Scalar} \vspace{0.02cm}&
\colhead{Vector} \\
\colhead{} &
\colhead{Lambert} &
\colhead{Isotropic} &
\colhead{Rayleigh} &
\colhead{Rayleigh} 
}
\startdata
0.005 & 0.0033334 & 0.00062685 & 0.0009394 & 0.00094 \\
0.01 & 0.0066668 & 0.0012574 & 0.0018826 & 0.00188 \\
0.05 & 0.033334 & 0.0064412 & 0.0095705 & 0.00963 \\
0.1 & 0.066668 & 0.013293 & 0.01956 & 0.0198 \\
0.15 & 0.1 & 0.020604 & 0.030015 & 0.03057 \\
0.2 & 0.13334 & 0.028429 & 0.040992 & 0.04201 \\
0.25 & 0.16667 & 0.036836 & 0.052557 & 0.0542 \\
0.3 & 0.2 & 0.045906 & 0.064788 & 0.06723 \\
0.35 & 0.23334 & 0.055736 & 0.077783 & 0.08121 \\
0.4 & 0.26667 & 0.066447 & 0.09166 & 0.09628 \\
0.45 & 0.3 & 0.078189 & 0.10657 & 0.11261 \\
0.5 & 0.33334 & 0.091154 & 0.12269 & 0.13041 \\
0.55 & 0.36667 & 0.10559 & 0.14029 & 0.14996 \\
0.6 & 0.40001 & 0.12183 & 0.15968 & 0.17161 \\
0.65 & 0.43334 & 0.14033 & 0.18131 & 0.19585 \\
0.7 & 0.46667 & 0.16174 & 0.20583 & 0.22336 \\
0.75 & 0.50001 & 0.18703 & 0.2342 & 0.25514 \\
0.8 & 0.53334 & 0.21777 & 0.26799 & 0.29281 \\
0.85 & 0.56668 & 0.25672 & 0.30994 & 0.33916 \\
0.9 & 0.60001 & 0.30957 & 0.36572 & 0.39991 \\
0.95 & 0.63334 & 0.392 & 0.45097 & 0.49069 \\
0.99 & 0.66001 & 0.53463 & 0.59585 & 0.64034 \\
0.999 & 0.66601 & 0.63582 & 0.69783 & 0.74337 \\
1.0 & 0.66668 & 0.68956 & 0.75165 & 0.79766 
\enddata
\tablecomments{These calculations were carried out with the formalism described in \citet{Madhu2012}.}
\label{deluxetable:Agtable}
\end{deluxetable}

The geometric albedo for a Lambert surface has a well-known analytic expression: $A_g(\omega) = (2/3)\,\omega$. Analytic fits to $A_g(\omega)$ for isotropic (iso) and scalar/vector (sca/vec) Rayleigh scattering are:
\begin{eqnarray}
A_g^\m{iso} &=& 0.6896 \frac{(1 + 0.4380 s) (1 - s)}{(1 + 1.2966 s)( 1 + 0.7269 s)}, \label{eqn:fit1} \\
A_g^\m{sca} &=& 0.8190 \frac{(1 + 0.1924 s) (1 - s)}{(1 + 1.5946 s)( 1 + 0.0886 \omega)}, \label{eqn:fit2} \\
A_g^\m{vec} &= & 0.6901 \frac{(1 + 0.4664 s) (1 - s)}{(1 + 1.7095 s)( 1 - 0.1345 \omega)}, \label{eqn:fit3}
\end{eqnarray}\\
where $s\equiv\sqrt{1 - \omega}$. These fits are accurate to within $0.5\%$ for all $\omega$, with the largest discrepancies corresponding to $\omega < 0.4$\footnote{\citet{Madhu2012} also provide an analytic fit to $A_g^\m{vec}(\omega)$, which is marginally better than Equation~(\ref{eqn:fit3}) for small $\omega$, but considerably less accurate for $\omega$ close to unity.}. Fits such as these are particularly useful for efficient theoretical calculations of the geometric albedo, as well as for retrieving a representative atmosphere-averaged scattering albedo from an observed geometric albedo. 

\begin{figure}[t!]
\centering
\includegraphics[width=9cm, trim=0.9cm 0cm 0.1cm 0cm,clip=true]{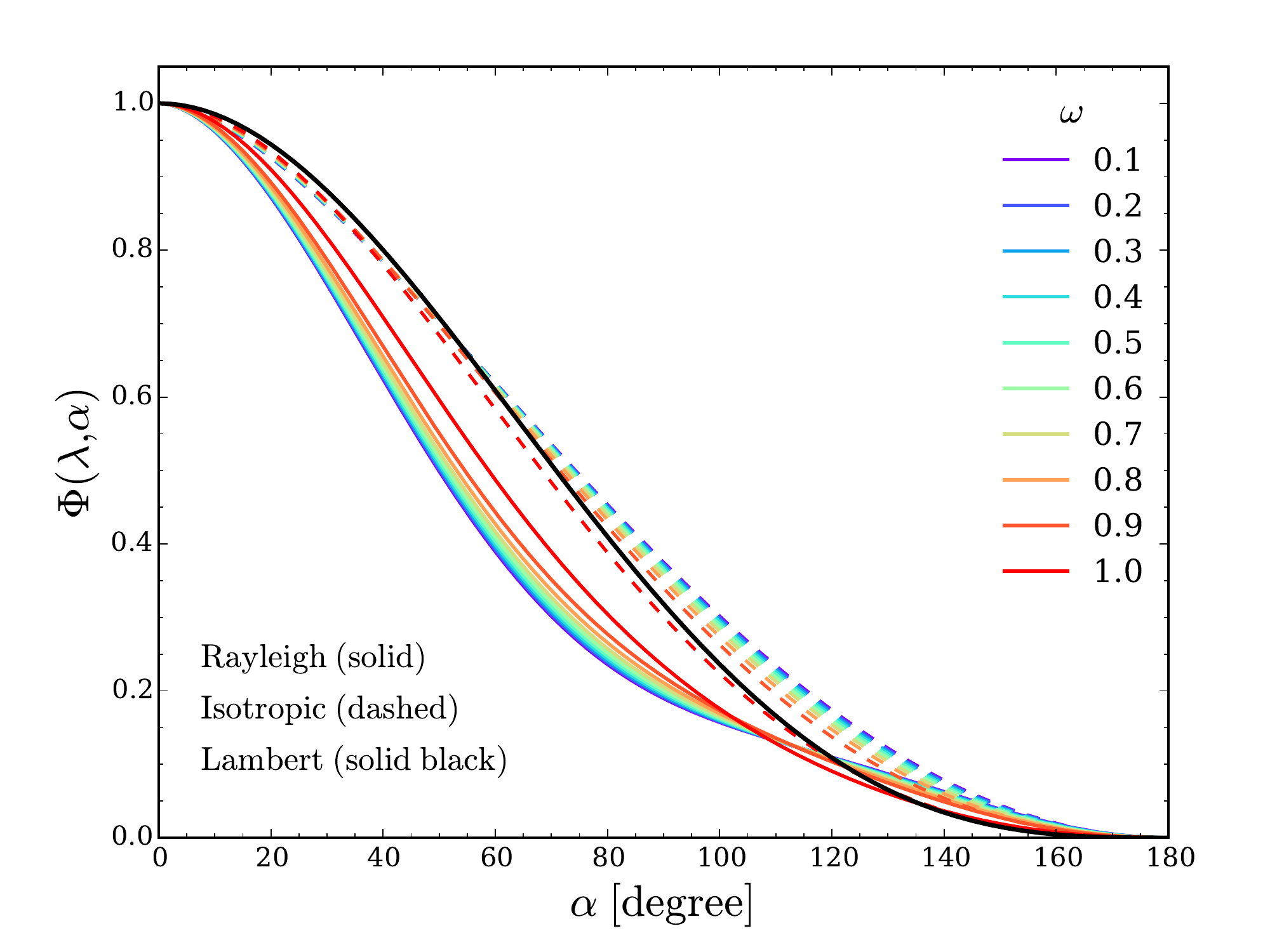}
\caption{Comparison of phase functions for Rayleigh, isotropic, and Lambert scattering. The  colors correspond to the indicated single-scattering albedos ($\omega$). Note that the Lambert phase function (solid black line) is independent of $\omega$.
\vspace{0.2cm}}
\label{fig:phase_curves}
\end{figure}
\begin{figure*}[t!]
\centering
\includegraphics[width=16cm, trim=1.cm 0.9cm 1.2cm 1cm,clip=true]{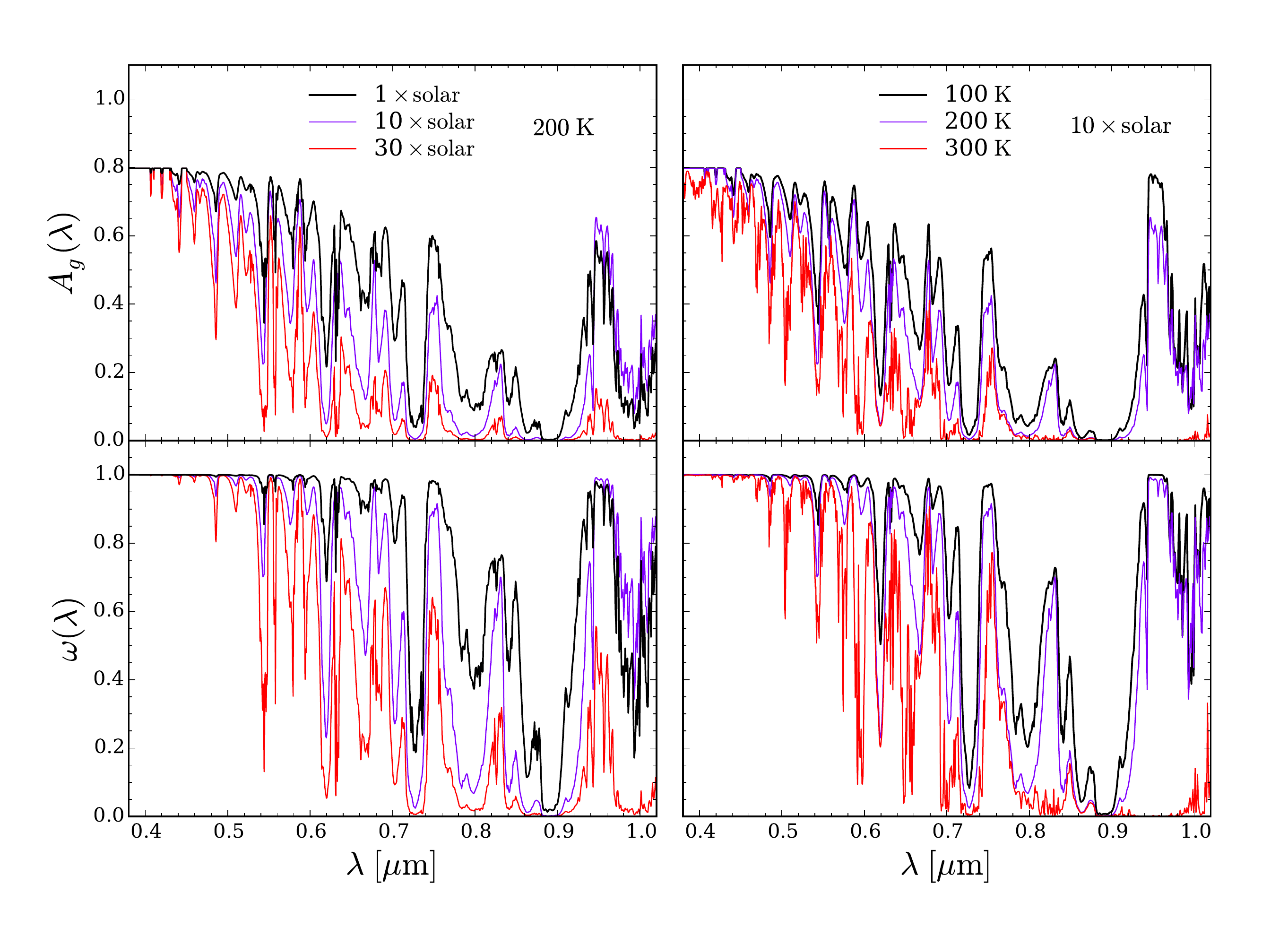}
\caption{Demonstration of the strong dependence of geometric (top row) and scattering (bottom row) albedo spectra upon temperature and atmospheric metallicity. The models in the left column fix the temperature at $T=200\ K$ and vary the metallicity, and those in the right column fix the metallicity at $10\,\times$\,solar and vary the temperature. Note the methane features near ${\sim}0.62~\m{\mu m}$, ${\sim}0.74~\m{\mu m}$, ${\sim}0.81~\m{\mu m}$, and ${\sim}0.89~\m{\mu m}$, ammonia spectral bands near ${\sim}0.65~\m{\mu m}$ and ${\sim}0.79~\m{\mu m}$, and broad water band at ${\sim}0.94~\m{\mu m}$. The models are cloud-free, homogeneous, semi-infinite, and use opacities at a pressure of 0.5 atmospheres. The calculations of $\omega(\lambda)$ use the opacity database from \citet{SharpBurrows2007} and the equilibrium chemical abundances from \citet{BurrowsSharp1999}. The geometric albedo spectra assume purely Rayleigh scattering atmospheres.}
\label{fig:scat_albedo}
\end{figure*}
\begin{figure}[t!]
\centering
\includegraphics[width=9cm, trim=0.6cm 0cm 0.8cm 0cm,clip=true]{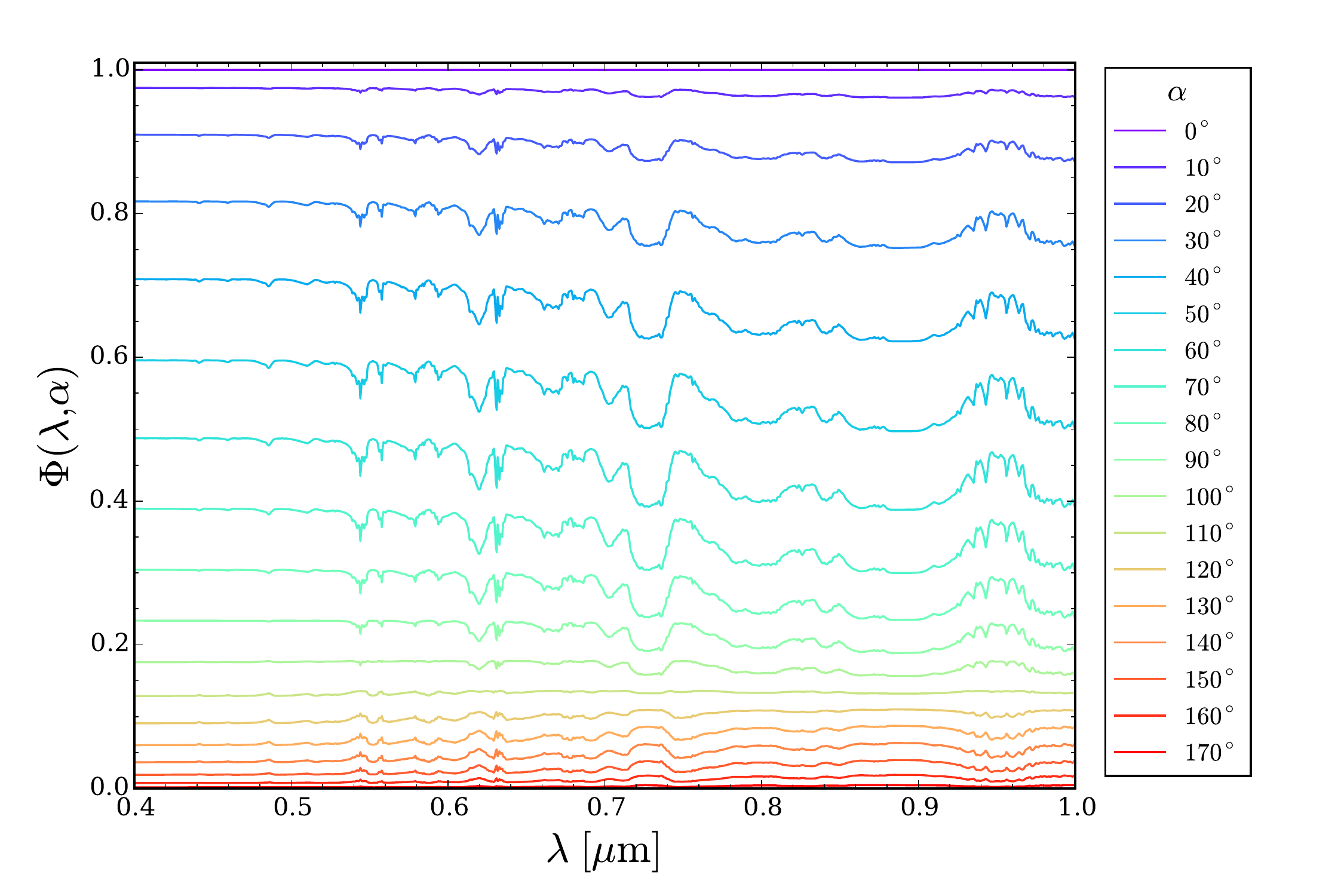}
\caption{Example variation with wavelength and phase angle of the Rayleigh phase function, where we have assumed the single-scattering albedo spectrum (shown as the black line in the bottom left panel of Figure~\ref{fig:scat_albedo}) of a semi-infinite homogeneous atmosphere with $T=200~K$, solar metallicity, and a pressure of 0.5 atmospheres. For a Lambert surface, the phase curve does not depend upon wavelength, and the horizontal lines corresponding to each phase would be flat and featureless, while of course the corresponding geometric albedos retain a stiff dependence upon wavelength.}
\label{fig:phi_lam}
\end{figure}

It is common to assume the Lambert phase function when modeling reflected-light observations. However, this assumption neglects the albedo- and wavelength-dependence of physical phase curves, and generally predicts higher values than the more physically motivated Rayleigh phase function \citep{Madhu2012}. Figure~\ref{fig:phase_curves} compares the phase curves for scattering due to Rayleigh, isotropic, and Lambert scattering. We see that Rayleigh scattering phase curves are systematically lower than isotropic and Lambert phase curves for $\alpha\lesssim120^\circ$, differing by as much as a factor of ${\sim}2$ for phase angles in the range $60^\circ\!\lesssim\alpha\lesssim\!90^\circ$ \citep[see also,][]{Madhu2012}. For most phase angles, increasing $\omega$ results in better agreement between these scattering processes. Rayleigh phase curves generally increase with increasing $\omega$, whereas isotropic scattering phase curves show the opposite trend. In \textsection\ref{sec:compare_scat}, we compare predictions of the direct detectability of EGPs with Rayleigh, isotopic, and Lambert phase functions. 

\subsubsection{Keplerian Elements}\label{sec:KE}
Once $\Phi(\lambda, \alpha)$ is obtained for a particular planet, the {\it observed} phase curve depends on the Keplerian elements of its orbit\footnote{See the appendix for a review of the equations that connect the Keplerian orbital elements to the observed phase angle and time.}. These include the semi-major axis ($a$), orbital inclination ($i$), eccentricity ($e$), argument of periastron ($\omega_p$), time of periastron passage ($t_p$), and longitude of the ascending node ($\Omega$) \citep{Sudarsky2005, Dyudina2005, Kane2011, Madhu2012}. For an assumed stellar mass, the period of the planet's orbit ($P$) can be calculated from $a$. Note $\Omega$ is necessary when an absolute celestial frame is designated (e.g., when measuring the full Stokes polarization or cataloguing many different systems in the same solar-system frame). However, one can assume $\Omega=90^\circ$ or simply ignore this parameter when the system itself is allowed to provide a natural orientation (e.g., in exoplanet radial-velocity measurements). Throughout this work, we will assume $\Omega=90^\circ$. 

\begin{figure*}[t!]
\centering
\includegraphics[width=16cm, trim=0.4cm 0.7cm 0.6cm 0.4cm,clip=true]{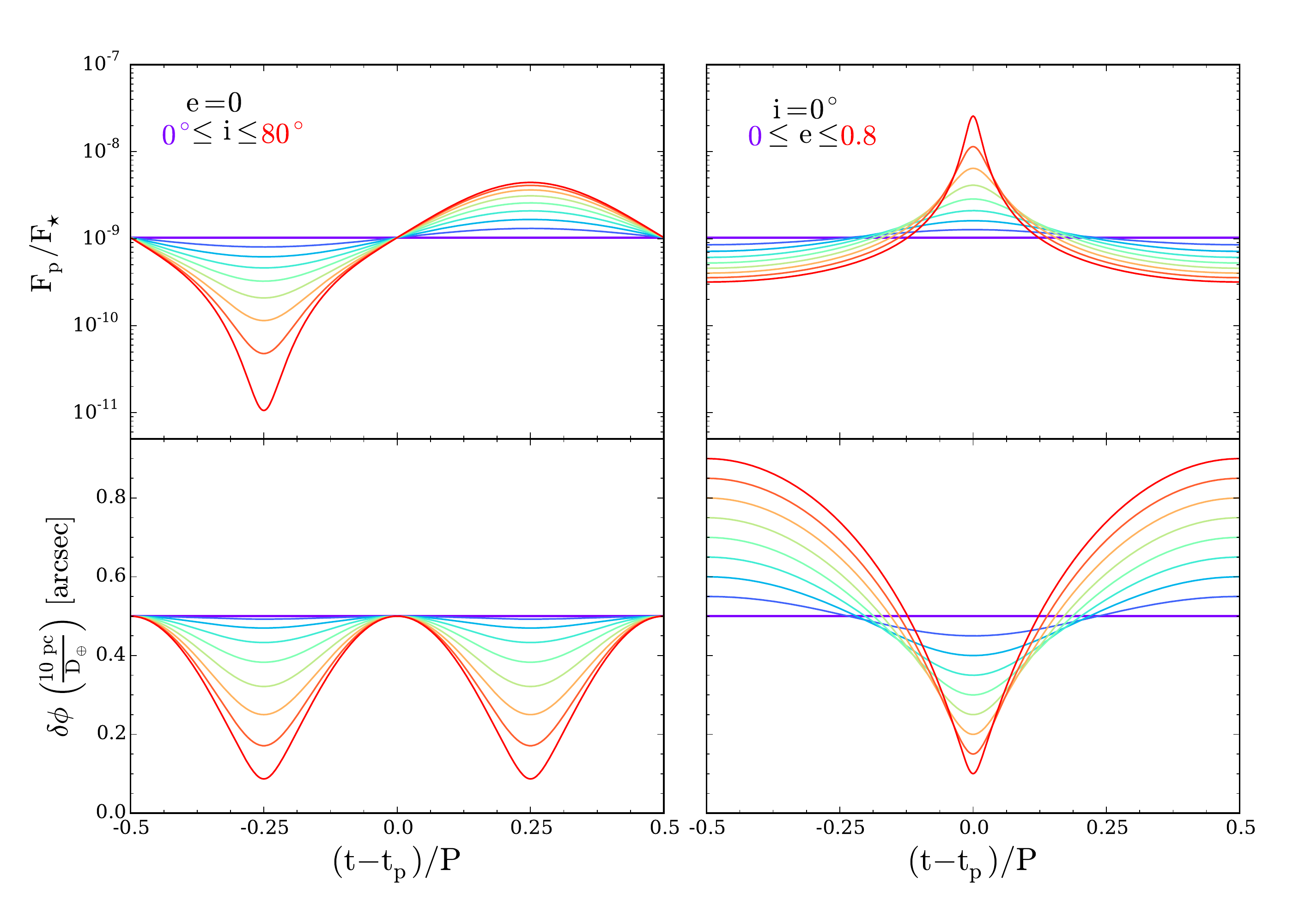}
\caption{The top row shows the time dependence of the planet/star flux ratio for circular orbits with many different inclinations (left) and face-on orbits with many different eccentricities (right). The bottom row shows the corresponding planet/star angular separation, assuming the system is located 10 pc from Earth. All calculations assume a Jupiter-like planet in an orbit with  $a = 5$~AU and $\omega_p = 0^{\circ}$.
\vspace{0.1cm}}
\label{fig:i-e_fr}
\end{figure*}

\subsubsection{Albedo Spectra from Homogeneous  Atmospheres}\label{sec:atms}
Semi-infinite homogeneous atmospheres are characterized by an average single-scattering albedo spectrum: $\omega(\lambda)\equiv\kappa_s/(\kappa_s + \kappa_a)$, where $\kappa_s$ is the scattering opacity, and $\kappa_a$ is the absorptive opacity. In this section, we provide illustrative scattering albedo spectra for cool EGPs, which will aid in the interpretation of some of the results presented in this paper. Our calculations use the opacity database from \citet{SharpBurrows2007} and the equilibrium chemical abundances from \citet{BurrowsSharp1999}. 

From its definition, it is clear that $\omega(\lambda)$ is a function of atmospheric parameters such as temperature, pressure, and composition. In Figure~\ref{fig:scat_albedo}, we demonstrate the characteristic strong dependence of scattering albedos (bottom row) and the corresponding Rayleigh scattering geometric albedos (top row) upon temperature and metallicity. The models are cloud-free, homogeneous, semi-infinite, and use opacities at a pressure of 0.5 atmospheres. For the temperatures and pressures relevant to cool EGPs, we find $\omega(\lambda)$ is relatively insensitive to variations in atmospheric pressure. In general, higher temperatures/metallicities result in lower albedos, particularly for wavelengths longward  of ${\sim}0.65~\m{\mu m}$. 

The spectra in Figure~\ref{fig:scat_albedo} show prominent methane features near ${\sim}0.62~\m{\mu m}$, ${\sim}0.74~\m{\mu m}$, ${\sim}0.81~\m{\mu m}$, and ${\sim}0.89~\m{\mu m}$, ammonia spectral bands near ${\sim}0.65~\m{\mu m}$ and ${\sim}0.79~\m{\mu m}$, and a broad water band at ${\sim}0.94~\m{\mu m}$. Although there is clearly some degeneracy between metallicity and temperature, this figure suggests optical flux ratio measurements, particularly between ${\sim}0.5-0.6~\m{\mu m}$ and/or ${\sim}0.82~\m{\mu m}$ and ${\sim}0.94~\m{\mu m}$, have great potential as temperature/composition diagnostics. 

Given a single-scattering albedo spectrum, one can use the formalism presented in \citet{Madhu2012} to calculate the phase function for an assumed scattering process as a function of phase angle and wavelength. Note that the often-assumed Lambert phase function is independent of $\omega(\lambda)$, and hence $\lambda$. In Figure~\ref{fig:phi_lam}, we show the phase function for Rayleigh scattering as a function of wavelength and phase angle, where we have assumed the scattering albedo spectrum shown as the black line in the bottom left panel of Figure~\ref{fig:scat_albedo}. As expected, the phase curve is equal to one for all $\lambda$ at $\alpha=0^\circ$, but note the variation in $\Phi(\lambda, \alpha)$ as $\alpha$ increases from $0^\circ$ to $170^\circ$. In contrast, a Lambert surface would result in featureless, horizontal lines for all $\alpha$. 

\subsection{Planet/Star Angular Separation}\label{sec:sep}
In addition to the planet/star contrast ratio, the planet/star angular separation is an essential parameter in determining the feasibility of the direct detection of reflected light from an exoplanet. For coronagraphic observations, the figure of merit is given as an IWA and OWA. An IWA (OWA) is the minimum (maximum) planet/star angular separation that can be resolved by a given instrument. Using the formalism outlined in the appendix, the projected angular separation, as a function of phase angle and time, can be obtained via the following equation:
\beq
\delta\phi = \frac{d}{D_\earth}\sqrt{\cos^2(\theta + \omega_p) + \sin^2(\theta + \omega_p)\cos^2 i}\,,
\eeq
where $D_\earth$ is the system's distance from Earth and $\theta$ is the planet's orbital angle as measured from periastron (the true anomaly). Inspection of the above equation reveals that, as with the planet/star flux ratio, the time dependence of the planet/star angular separation will depend sensitively upon the Keplerian orbital elements.

\begin{figure*}[t!]
\centering
\includegraphics[width=18.05cm,trim=0.2cm 0.cm 0.2cm 0cm,clip=true]{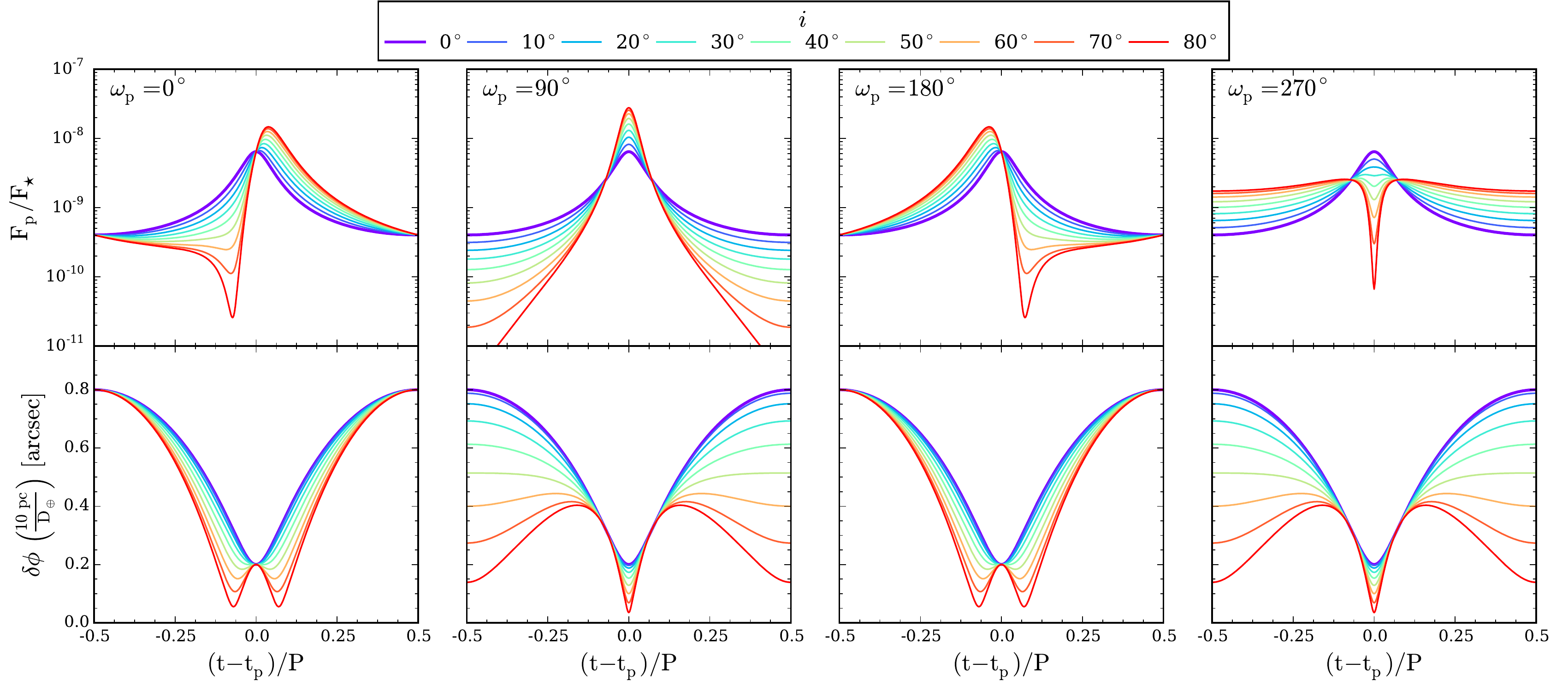}
\caption{Planet/star flux ratios (top row) and angular separations (bottom row) for a Jupiter-like planet at a distance $D_\oplus=10$~pc from Earth on an orbit with $e=0.6$ and $a=5$~AU. Each column corresponds to orbits with the periastron argument indicated in each top left corner, and the colors represent the indicated inclinations. For the orbit orientation, we use the convention $\alpha = \theta = 0^\circ$ when $\omega_p= i = 90^\circ$, implying full phase corresponds to the time of periastron passage when $\omega_p = 90^\circ$. Note that the detectability of planets on face-on orbits (purple lines) is independent of $\omega_p$.
\vspace{0.1cm}}
\label{fig:i-wp_fr}
\end{figure*}

\section{Variation of Flux Ratios and Angular Separations with Orbital Parameters}\label{sec:fr_KE}
Using the methods presented in \textsection\ref{sec:detect} to calculate planet/star flux ratios and angular separations, we now present example realizations of the $(F_p/F_\star) - \delta\phi$ parameter space for EGPs. The calculations in this section are representative of those that will be used in our Monte Carlo experiments, which are presented in \textsection\ref{sec:MC}. Several previous studies have also investigated this parameter space. In particular, \citet{Kane2010,Kane2011} considered the eccentricity and inclination dependence of exoplanet phase signatures, \citet{Kane2013} used the eccentricity distribution of known exoplanets to study the completeness of imaging surveys, and \citet{Janson2010} studied the effect that prior knowledge of orbital inclination has on the efficiency of directly imaging Earth-like planets. In this paper, we build on previous work by performing a systematic study of the dependence of the direct detectability of EGPs upon a wide range of technological and astrophysical parameters (\textsection\ref{sec:results}). 

\subsection{Eccentricity and Inclination}
As mentioned in \textsection\ref{sec:KE} and \textsection\ref{sec:sep}, observed flux ratios and separations are strong functions of the Keplerian orbital elements. Figure~\ref{fig:i-e_fr} demonstrates this dependence for circular orbits with many different inclinations (left column) and face-on orbits with many different eccentricities (right column). The top row shows planet/star flux ratios, and the bottom row shows the corresponding angular separations for a system located 10~pc from Earth. All calculations in this figure assume a Jupiter-like planet in an orbit with $a=5$~AU and $\omega_p=0^\circ$. 

Although it has been pointed out previously in the literature, it is worth reemphasizing that direct-imaging observations are well-suited for studying face-on systems (the opposite is true for the radial-velocity and transit methods, which are biased towards high-inclination systems). The reason for this is that an exoplanet on a circular, face-on orbit with $F_p/F_\star>C_\m{min}$ and $\delta\phi > $~IWA will be directly detectable at all times (note we are assuming $\delta\phi < $~OWA ). If, on the other hand, its inclination is ${\sim}90^\circ$, both the planet/star flux ratio and angular separation will drop below their detection thresholds for some fraction of the orbit as the planet passes through the full range of possible phase angles. For instance, if $C_\m{min} = 10^{-10}$ and $\m{IWA}=0.3''$, then a face-on orbit (flat purple lines in Figure~\ref{fig:i-e_fr}) will be well above the detection thresholds at all times. However, as we increase the inclination in the left column of this figure, we see that the flux ratio dips below $C_\m{min}$ for an increasingly larger fraction of the orbit centered  at new phase ($\alpha=180^\circ$). Similarly, the minimum planet/star angular separation decreases with increasing inclination, falling below the IWA for an increasingly larger fraction of the orbit centered at new phase {\it and} full phase. Although the flux ratio near full phase is largest for high-inclination orbits, the planet will still be undetectable due to the small planet/star angular separation. 

In the right column of Figure~\ref{fig:i-e_fr}, we show the effect of increasing the eccentricity of a face-on orbit from circular ($e=0$) to highly eccentric ($e=0.8$). In this case, the phase variations are due to the time dependence of the orbital distance. On one hand, a planet on a highly eccentric orbit can be an excellent direct-imaging target, as it spends a large fraction of its orbit at larger orbital distances from its parent star, and for certain sets of Keplerian elements, this can provide angular separations that are highly favorable for direct-imaging observations. This, however, comes at the cost of lower flux ratios throughout most of its orbit and very small angular separations when it is near periastron passage. 

\subsection{Argument of Periastron and Inclination}

As has been pointed out by \citet{Kane2011}, the interaction between the argument of periastron and inclination in $(F_p/F_\star) - \delta\phi$ parameter space can be quite complex. Figure~\ref{fig:i-wp_fr} shows various combinations of these parameters for a Jupiter-like planet on an eccentric orbit with $e=0.6$ and $a=5$~AU. Each column shows planet/star flux ratios and angular separations for orbits with inclinations that vary from $0^\circ$ to $80^\circ$, with the argument of periastron held constant at the value indicated in each top left corner. Note we use the convention that when $\alpha = \theta = 0^\circ$, $\omega_p= i = 90^\circ$. In other words, periastron passage coincides with full phase when the argument of periastron is $90^\circ$ and with new phase when it is $270^\circ$.

The detectability of planets on face-on orbits is independent of $\omega_p$, as can be seen by the constant purple curve in all panels of Figure~\ref{fig:i-wp_fr}. As we increase the inclination, the flux ratio can either increase or decrease, depending upon the periastron argument and phase of the orbit. For $\omega_p=90^\circ$, the flux ratio decreases with inclination for most of the orbit as the planet transitions from being permanently oriented at quarter phase when $i=0^\circ$ to displaying nearly all phase angles when $i=80^\circ$.  At full phase, which corresponds to periastron passage, the flux ratio achieves its maximum for this set of parameters. However, as is typically the case, this maximum flux ratio corresponds to a minimum planet/star angular separation that, in the limit of a perfectly edge-on system, drops to zero at the time of periastron passage. Interesting behavior also occurs when $\omega_p = 270^\circ$. For this periastron argument, periastron passage corresponds to new phase, and, in exactly the opposite sense from the $\omega_p=90^\circ$ case, the flux ratio increases with inclination for most of the orbit and results in a minimum at the time of periastron passage for high-inclination orbits. For orbits with $i\gtrsim50^\circ$, the minimum angular separation actually coincides with the minimum flux ratio, and the maximum angular separation coincides with the relatively large fraction of the orbit with a large flux ratio. 

\section{Monte Carlo Experiments}\label{sec:MC}
\begin{figure}[t!]
\centering
\includegraphics[width=9cm, trim=0.5cm -0.1cm 0.cm 0.2cm,clip=true]{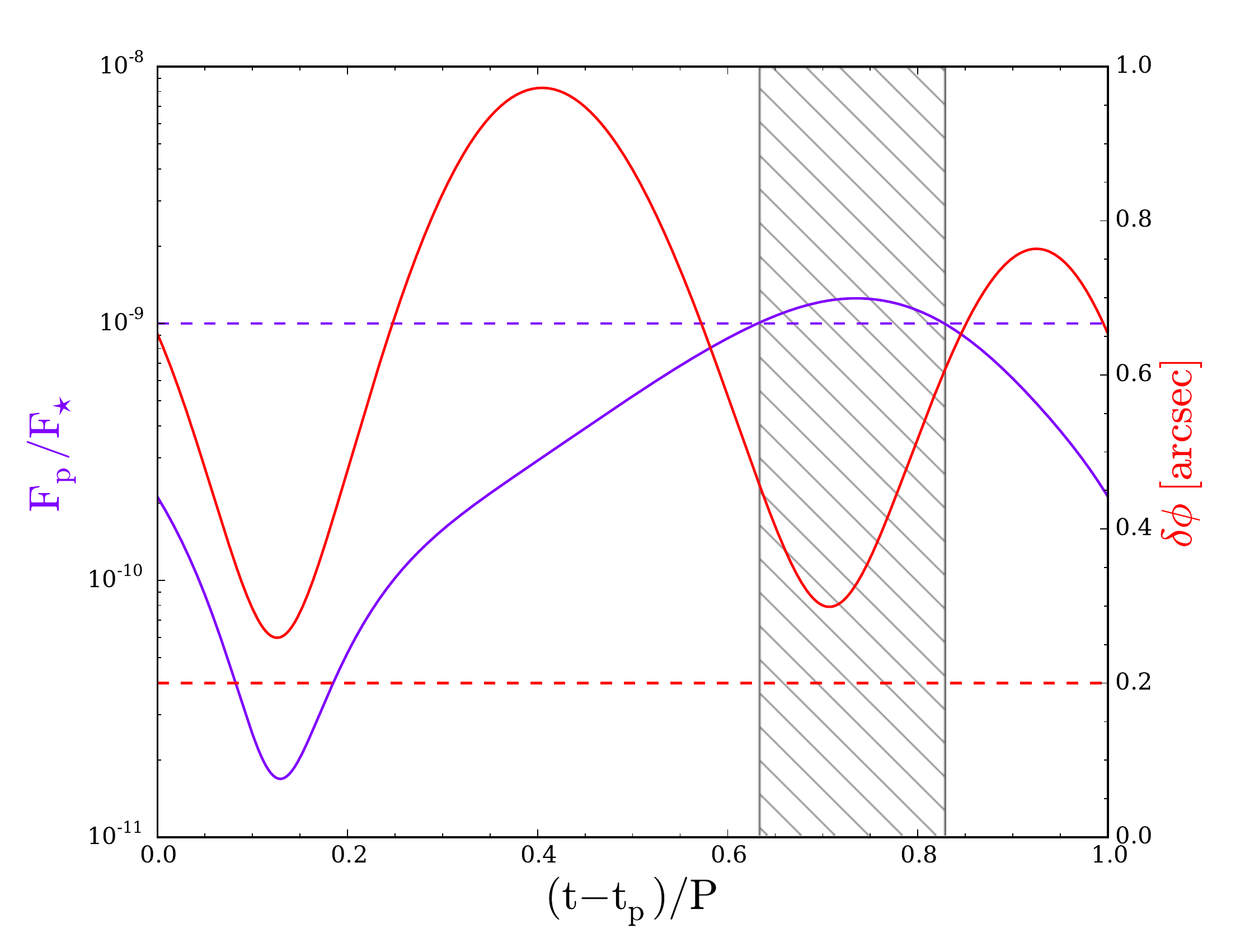}
\caption{The planet/star flux ratio (purple, left axis) and angular separation (red, right axis) of a Jupiter-like planet orbiting a star 10~pc from Earth on a randomly generated orbit with $\{a,\ e,\ i,\ \omega_p\}=\{8.7~\m{AU},\ 0.14,\ 71^\circ,\ 211^\circ\}$. Assuming $C_\m{min} = 10^{-9}$ and $\m{IWA} = 0.2''$, which are indicated by the purple and red dashed lines, this planet would be directly detectable at times within the gray hatched region. The observability fraction of this planet is $f_\m{obs}\approx20\%$.}
\label{fig:observe_frac}
\end{figure}

We have performed a set of Monte Carlo (MC) experiments, in which we generate random orbits and calculate the fraction of time a giant planet on each orbit would be directly detectable in the optical. We define this fraction to be the observability fraction, $f_\m{obs}$, and we calculate its value under many different assumptions. An illustrative calculation of $f_\m{obs}$ for a Jupiter-like planet on a random orbit around a star located 10~pc from Earth is shown in Figure~\ref{fig:observe_frac}. We have assumed a direct-imaging observatory with $C_\m{min} = 10^{-9}$ and $\m{IWA} = 0.2''$, which are the baseline coronagraph design parameters for WFIRST/AFTA. The purple curve represents the planet/star contrast ratio (left axis), the red curve shows the corresponding angular separation (right axis), and the corresponding detection thresholds are indicated by the dashed lines. For the planet to be detectable, each curve must be above its associated dashed line. This is true only for times within the gray hatched region. A Jupiter-like planet on this particular orbit would have $f_\m{obs}\approx20\%$. Given an estimate of the occurrence rate of such planets, the mean $f_\m{obs}$ for many systems can be translated into the probability that a direct-imaging observation of a random star will result in a detection (see \textsection\ref{sec:probs}). 

\begin{figure*}[t!]
\centering
\includegraphics[width=16.cm, trim=1.4cm 1.55cm 1.4cm 0.3cm,clip=true]{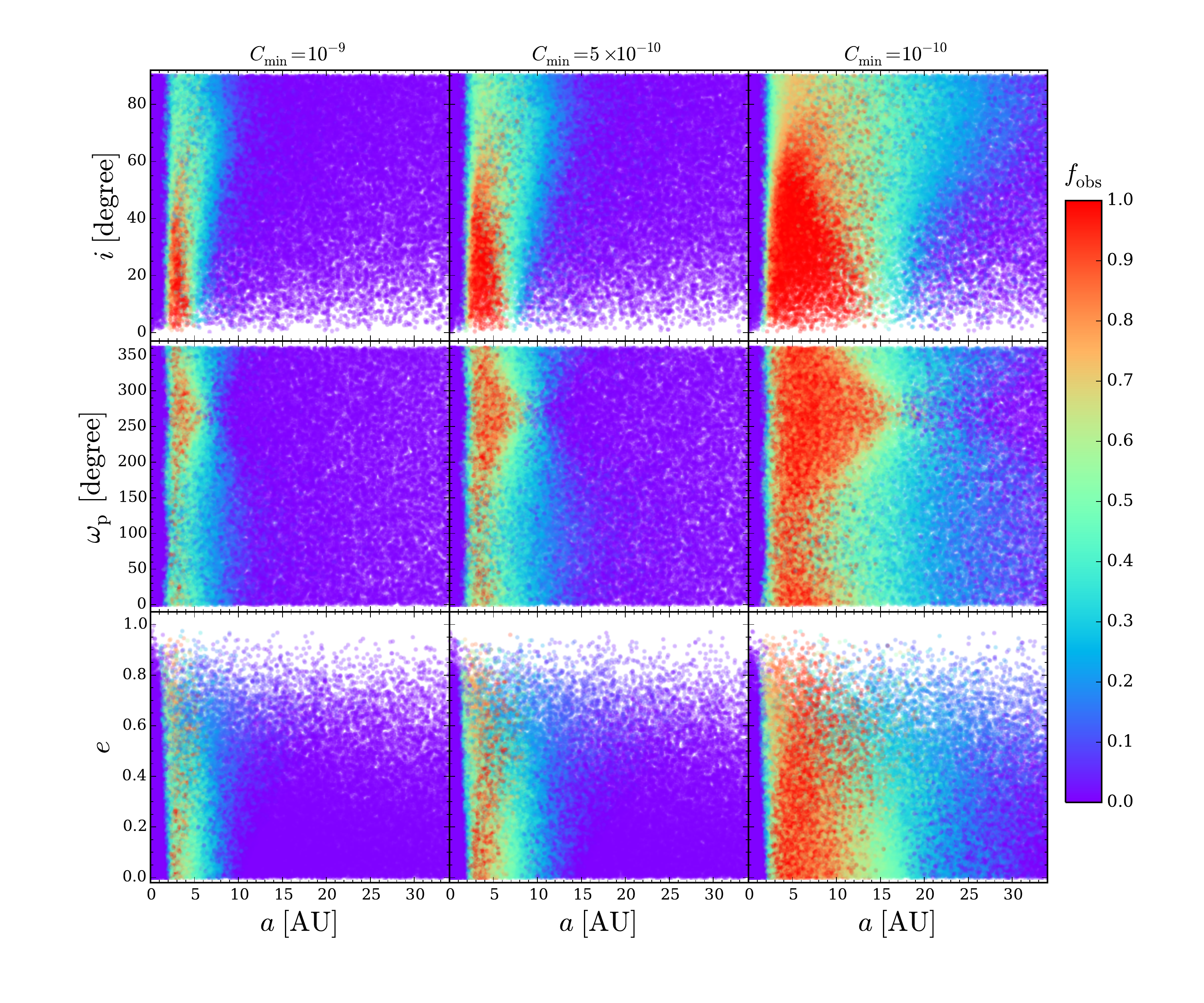}
\caption{Distribution of observability fractions ($f_\m{obs}$) as a function of orbital parameters for three illustrative MC experiments, where we have assumed $C_\m{min}=10^{-9}$ (left column), $C_\m{min}=5\times10^{-10}$ (middle column), and $C_\m{min}=10^{-10}$ (right column). Each experiment is composed $5\times10^5$ random orbits generated from the distributions given in Table~\ref{deluxetable:distributions}. In each panel, dots indicate a single orbit, and the color-scale represents $f_\m{obs}$, which we calculate for direct observations of a Jupiter-like planet orbiting a star at a distance of $D_\oplus=10$~pc from Earth, using an instrument with $\m{IWA}=0.2''$, $\m{OWA}=1.75''$, and the minimum contrast indicated at the top of each column.}
\label{fig:mc_params}
\end{figure*}

\subsection{Orbital Parameter Distributions}\label{sec:param_dist}
Each random orbit requires the generation of four Keplerian elements: $\{a,\ e,\ i,\ \omega_p\}$. To accomplish this, we assume planetary orbital orientations are isotropically distributed with respect to the observer, which leads to periastron arguments distributed uniformly in $[0,\ 2\pi]$ and $\cos i$ distributed uniformly in $[-1,1]$. The semi-major axis and eccentricity distributions must be determined empirically. Most of our knowledge of these distributions has come from radial-velocity \citep{Cumming2008, Howard2010} and transit \citep{Batalha2013, Fressin2013} surveys that are biased towards short-period planets and have only been in operation long enough to fully describe planets with periods up to a few thousand days. 

Nevertheless, we assume both the semi-major axis and eccentricity distributions derived from radial-velocity planets can be extrapolated to wider separations. For the semi-major axis distribution, we adopt a power-law, $dN/da \propto a^{\beta}$, where our fiducial calculations assume $\beta=-0.61$ \citep{Cumming2008}. This distribution is based on radial-velocity planets with orbital periods in the range $2 - 2000$~days, corresponding to semi-major axes in the range $0.03-3.0~\m{AU}$. Results from direct-imaging surveys have been used to set model-dependent limits on the maximum semi-major axis ($a_\m{max}$) to which the radial-velocity distribution can be extrapolated. Using Monte Carlo simulations to examine the null results from direct-imaging searches for EGPs around 118 stars, \citet{Nielsen2010} determined $a_\m{max}\!\sim\!65-234$~AU, depending on the assumed planet luminosity model.  \citet{Brandt2014} performed a statistical analysis of direct-imaging data from a sample of nearly 250 stars, finding $a_\m{max}\!\sim\!30-100$~AU, which again depends on the adopted cooling model. With these studies as our motivation, our fiducial calculations assume the semi-major axis distribution is given by a power-law with a minimum $a_\m{min}=0.03$~AU and a maximum $a_\m{max}=65$~AU.

For the eccentricity distribution, we use the two-parameter Beta distribution given by \citet{Kipping2013}:
\beq\label{eqn:beta_dist}
P_\beta(e;a,b) = \frac{\Gamma(a+b)}{\Gamma(a)\Gamma(b)}e^{a-1}(1-e)^{b-1},
\eeq
where $\Gamma$ is the Gamma function. Our fiducial calculations assume $a=1.12$ and $b=3.09$, which are the long-period parameters given by \citet{Kipping2013}, who derived this distribution by regressing the known cumulative density function of orbital eccentricities from exoplanets detected from the radial-velocity method. The orbits of planets with $a<0.1$~AU are expected to be rapidly circularized due to strong tidal forces. Although we do not expect the eccentricities of such planets to influence our results, we set $e=0$ for any planet with $a<0.1$~AU. 

We find the direct detectability of EGPs is highly sensitive to the assumed semi-major axis distribution, but relatively insensitive to the eccentricity distribution. Our fiducial orbital parameter distributions are summarized in Table~\ref{deluxetable:distributions}.

\begin{deluxetable}{ccc}[h!]
\centering
\tablecaption{Fiducial Orbital Parameter Distributions}
\tablewidth{8.3cm}
\tablecolumns{3}
\tablehead{
\colhead{Paramter}&
\colhead{Range}&
\colhead{Distribution}
}
\startdata
$a$ 				  &   [0.03  AU, 65 AU]    &   $dN/da \propto a^\beta$\\[0.05cm]
$e$                           & $[0,1)$          &   $P_\beta(e; a, b)$\footnote{This is the Beta distribution from \citet{Kipping2013} (Equation~(\ref{eqn:beta_dist})).}\\[0.05cm]
$\cos i$							&  [-1, 1]  &  uniform\\[0.05cm]
$\omega_p$   &  [0, $2\pi$]   &  uniform
\enddata
\tablecomments{For our fiducial semi-major axis distribution, we assume a power-law index $\beta=-0.61$ \citep{Cumming2008}. For our fiducial eccentricity distribution, we assume $a=1.12$ and $b=3.09$, which are the long-period parameters given in \citet{Kipping2013}. We set $e=0$ for any planet with $a<0.1$~AU.}
\label{deluxetable:distributions}
\end{deluxetable}

\section{Results}\label{sec:results}
\subsection{Mean Observability Fractions}\label{sec:mean_fobs}

Here, we present a parametric study of the mean observability fraction, \fobs, for different sets of MC experiments, each composed of $5\times10^5$ random orbits. For a given population of planets, \fobs\ can be interpreted as a detection probability, assuming all stars have one such planet. In Figure~\ref{fig:mc_params}, we show the distribution of $f_\m{obs}$ as a function of orbital parameters for three illustrative MC experiments. Each column corresponds to an independent experiment, where we have assumed $C_\m{min}=10^{-9}$ (left column), $C_\m{min}=5\times10^{-10}$ (middle column), and $C_\m{min}=10^{-10}$ (right column). In all three cases, we use our fiducial orbital parameter distributions, and calculate $f_\m{obs}$ for direct observations of a Jupiter-like planet orbiting a star at a distance of $D_\oplus=10$~pc from Earth, using an instrument with $\m{IWA}=0.2''$, $\m{OWA}=1.75''$, and the minimum contrast specified at the top of each column. 

In a given panel, each dot represents a single orbit, and the color-scale shows the observability fraction for the indicated assumptions about the planet and direct-imaging observatory. For a fixed minimum contrast, the semi-major axis has by far the largest effect on $f_\m{obs}$, creating highly clustered regions of low and high values of $f_\m{obs}$. The sharp drop in observability interior to 2~AU corresponds to an angular separation of $0.2''$, which is the assumed IWA. Note, however, that eccentric orbits with $a<2$~AU are able to achieve nonzero observability fractions, since a planet on such an orbit spends a large fraction of its time near orbital distances of $a(1+e)$, making it possible for the observed angular separations to fall outside the IWA, provided the full set of Keplerian elements allows for such geometries. These results make it clear that $f_\m{obs}$ is very sensitive to the assumed semi-major axis distribution. Steeper (shallower) power-law indices will result in narrower (broader) regions of high $f_\m{obs}$ values, and the assumed values of $a_\m{max}$ and $a_\m{min}$ determine the region within which the $5\times10^5$ planets are distributed. In addition, decreasing the minimum contrast increases the range of semi-major axes that are detectable and leads to an increasing number of orbits for which $f_\m{obs}\!\sim\!1.0$ (i.e., orbits that are observable at all times). For $C_\m{min}=10^{-9}$, low- to modest-eccentricity orbits with $f_\m{obs}>1\%$ are confined to $2~\m{AU}\lesssim a \lesssim 10~\m{AU}$. As we decrease $C_\m{min}$ from left to right in this figure, we see orbits with increasingly larger semi-major axes achieve nonzero $f_\m{obs}$, reaching $a\sim15$~AU and $a\sim35$~AU for $C_\m{min}=5\times10^{-10}$ and $C_\m{min}=10^{-10}$, respectively.

As mentioned previously, low-inclination orbits are beneficial to direct-imaging observations. In the top left panel, it is particularly clear that orbits with $f_\m{obs}\!\sim\!1.0$ are associated with low inclinations. However, planets with $i\gtrsim50^\circ$ can have $f_\m{obs}\!\sim\!15\%$ out to $a\!\sim\!10 - 30$~AU, depending on $C_\m{min}$, since phase angles near full phase become observationally accessible. There is also an interesting feature near $\omega_p\!\sim\!270^\circ$, which corresponds to orbits with periastron passage occurring at new phase. Near this periastron argument, the region of large $f_\m{obs}$ extends to semi-major axes as large as ${\sim}20$~AU for $C_\m{min}=10^{-10}$. For a planet with $\omega_p\!\sim\!270^\circ$, the planet/star flux ratio rises with increasing inclination throughout most of its orbit, and the largest angular separations coincide with the highest flux ratios when $i\gtrsim50^\circ$ (see Figure~\ref{fig:i-wp_fr}). These factors collectively increase the semi-major axes for which a planet on such an orbit can have a large observability fraction.

For the remainder of this section, we present the mean observability fraction, \fobs, for many MC experiments, which are run under different sets of assumptions. In each case, our goal is to study how the observability fraction depends upon input parameters such as albedo, planetary radius, distance from Earth, minimum achievable contrast, and the semi-major axis distribution. For each experiment, we vary one parameter with respect to a fiducial model, which is comprised of the following assumptions:
\begin{itemize}
\item the planetary atmospheres are Rayleigh scattering with $A_g = 0.5$ (this corresponds to $\omega\approx0.95$ for Rayleigh scattering),
\item the planets have the radius of Jupiter and orbit stars located 10~pc from Earth,
\item the direct-imaging observatory has an $\m{OWA}=1.75''$ and $C_\m{min}=10^{-9}$ (similar to what is anticipated for WFIRST/AFTA),
\item the Keplerian orbital elements are distributed according to the distributions given in Table~\ref{deluxetable:distributions}.
\end{itemize}

Of course, the above assumptions do not fully represent realistic physical/statistical properties of planets or any future direct-imaging mission. Nevertheless, by varying each parameter of this very simple reference model, it is possible to gain insight into the general problem of directly detecting giant exoplanets in the optical, both in the case of a blind search and observations of stars known to host giant planets. 

\begin{figure*}[t!]
\centering
\includegraphics[width=19cm, trim=2.65cm 0.1cm 0.5cm 0.1cm,clip=true]{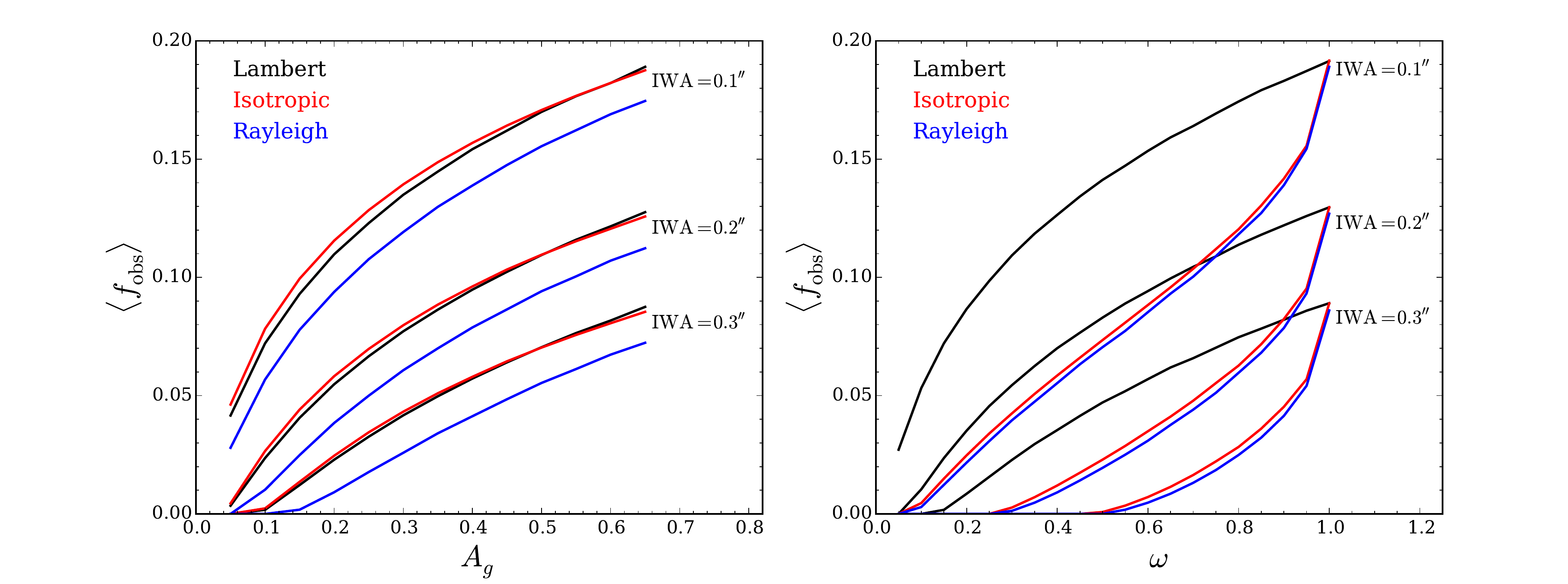}
\caption{ The dependence of the mean observability fraction \fobs\ upon the geometric albedo ($A_g$, left) and single-scattering albedo ($\omega$, right), assuming Lambert (black), isotropic (red), and Rayleigh (blue) scattering atmospheres. In the left panel, we assume $A_g$ and interpolate Table~\ref{deluxetable:Agtable} to infer the corresponding $\omega$ for each scattering process. In contrast, in the right panel, we assume $\omega$ and infer $A_g$ for each scattering process. For each calculation, we assume Jupiter-size planets that orbit stars located 10 pc from Earth, with random orbital parameters drawn from our fiducial distributions (Table~\ref{deluxetable:distributions}). The direct-imaging observatory is assumed to have an $\m{OWA}=1.75''$ and $C_\m{min}=10^{-9}$. The three sets of curves in both panels correspond to the indicated IWAs. The WFIRST/AFTA coronagraph is anticipated to have $\m{IWA=0.2''}$.}
\label{fig:fix_Ag_om}
\end{figure*}

\subsubsection{Dependence on Scattering Properties}\label{sec:compare_scat}
We start by varying the first of our fiducial model assumptions. Figure~\ref{fig:fix_Ag_om} shows the dependence of \fobs\ upon the geometric albedo (left panel) and the single-scattering albedo (right panel), assuming Lambert (black), isotropic (red), and Rayleigh (blue) scattering atmospheres. Each curve is composed of many distinct MC experiments, and the three sets of curves in each panel correspond to the indicated IWAs. The WFIRST/AFTA coronagraph is anticipated to have $\m{IWA}\sim0.2''$. Note this value will depend upon the angular separation and wavelength of the observation. For each scattering mechanism, there is a one-to-one correspondence between $\omega$ and $A_g$ (see Table~\ref{deluxetable:Agtable}). Additionally, the phase curves for isotropic and Rayleigh scattering are direct functions of $\omega$, and the Lambert phase curve is independent of $\omega$ (see Figure~\ref{fig:phase_curves}). Thus, given $A_g$ or $\omega$, it is straightforward to calculate the product $A_g\cdot\Phi(\alpha)$, which is proportional to the planet/star flux ratio, for an assumed scattering or effective scattering mechanism.  

In the left panel of Figure~\ref{fig:fix_Ag_om}, we show the dependence of \fobs\ upon $A_g$. For each MC experiment, we assume $A_g$ and interpolate Table~\ref{deluxetable:Agtable} to infer the corresponding $\omega$, which is needed to calculate the phase curves for Rayleigh and isotropic scattering. For a given $A_g$, the scattering phase functions are entirely responsible for the differences seen between the assumed scattering processes. Note that, since the Lambert phase function is independent of $\omega$, the phase function is the same for each black curve. As can be seen in Figure~\ref{fig:phase_curves}, Rayleigh scattering phase curves are systematically lower than isotropic and Lambert phase curves for $\alpha\lesssim120^\circ$. It is for this reason that \fobs\ is always the lowest for Rayleigh scattering. For most $A_g$, Rayleigh scattering predicts observability fractions that are ${\sim}20\%$ below the predictions of isotropic and Lambert scatting, regardless of the IWA.  At the anticipated IWA for WFIRST/AFTA, \fobs\ decreases sublinearly with $A_g$ for each scattering process, with maximum and minimum values for Rayleigh scattering of ${\sim}11\%$ for $A_g=0.65$ and ${\sim}1\%$ for $A_g \lesssim 0.1$.  For context, Figure~\ref{fig:scat_albedo} suggests that cool EGPs may have large geometric albedos ($A_g\!\sim\!0.6-0.8$) in the optical, particularly for wavelengths $\lesssim0.6~\mu$m and near the broad water band at ${\sim}0.94~\mu$m. As expected, these results suggest planetary geometric albedos can have a dramatic effect on the results of an optical direct-imaging survey.

\begin{figure}[t!]
\centering
\includegraphics[width=9.2cm, trim=0.3cm 0.cm -0.2cm 0cm,clip=true]{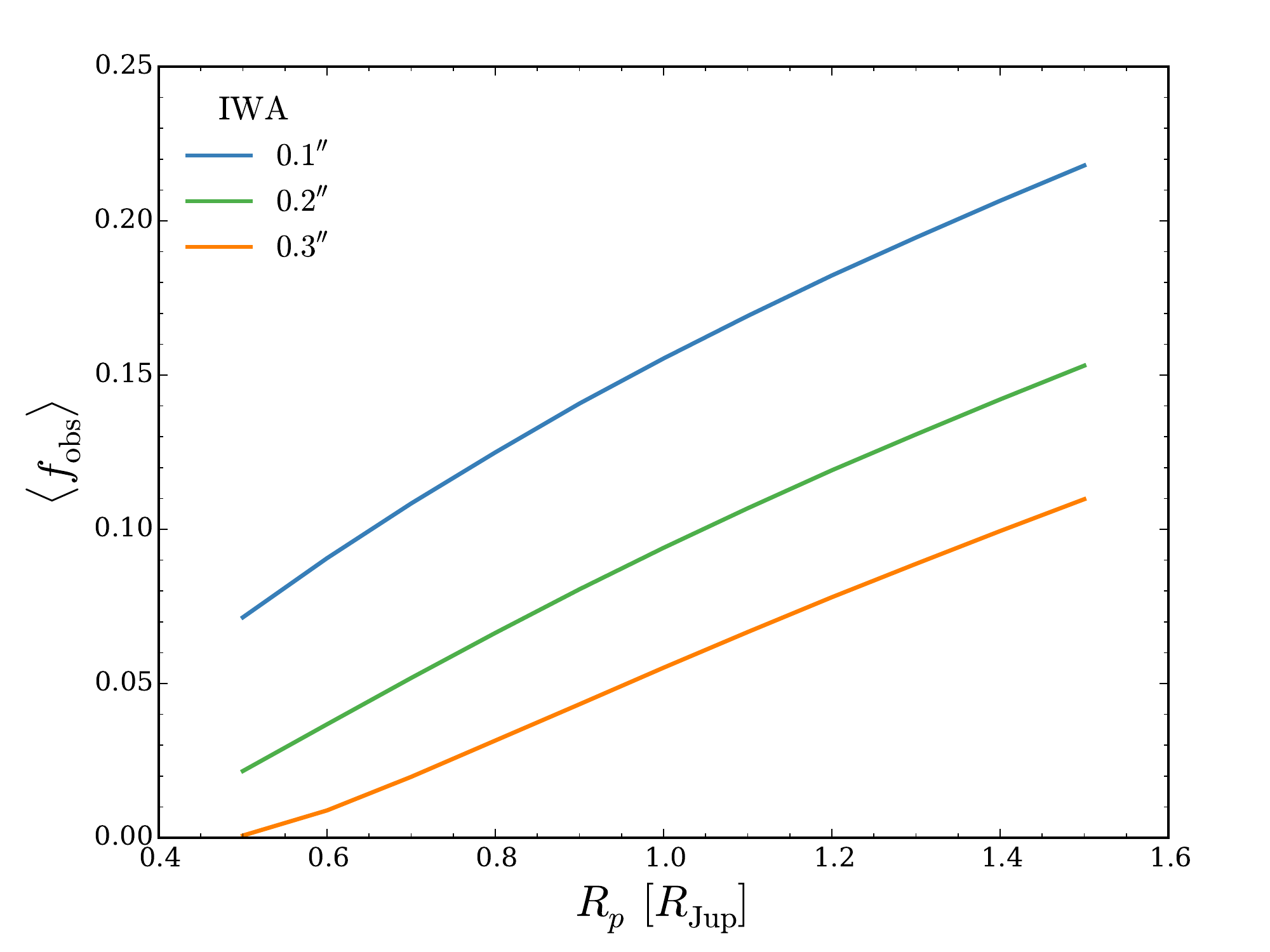}
\caption{Variation of the mean observability fraction \fobs\ with planet radius. With the exception of planet radius, all assumptions are as listed at the end of \textsection\ref{sec:mean_fobs}. The three curves show \fobs\ for observations made with a coronagraph with the specified IWAs. We note that gas giants are expected to have $0.8~R_\m{Jup} \lesssim R_p \lesssim 1.2~R_\mathrm{Jup}$, but we show a wider range of radii for illustrative purposes.}
\label{fig:vary_Rp}
\end{figure}

In the right panel of Figure~\ref{fig:fix_Ag_om}, we show the dependence of \fobs\ upon $\omega$. In contrast to the left panel, here we assume $\omega$ for each MC experiment and use Table~\ref{deluxetable:Agtable} to infer the corresponding $A_g$. In this case, both $A_g$ and $\Phi(\alpha)$ contribute to the differences  between the assumed scattering processes. We see that \fobs\ is comparable for Rayleigh and isotropic scattering for all $\omega$. This arises from the fact that, for a given $\omega$, $A_g$ for Rayleigh scattering is systematically {\it higher} than that for isotropic scattering, whereas $\Phi(\alpha)$ for Rayleigh scattering is systematically {\it lower}, leading to roughly consistent observability fractions. For conservative scattering ($\omega=1.0$), the geometric albedos for Lambert, isotropic, and Rayleigh scattering differ by less than ${\sim}13\%$, and  \fobs\ is essentially independent of the assumed scattering mechanism. However, for non-conservative scattering ($\omega < 1.0$), the geometric albedos for isotropic and Rayleigh scattering can be less than those for Lambert scattering by as much as a factor of ${\sim}4$. As a consequence, Lambert scattering generally predicts significantly higher \fobs\ than both isotropic and Rayleigh scattering.

As a specific example, consider $\omega = 0.6$. For this scattering albedo, the Lambert phase function corresponds to a geometric albedo that is a factor of ${\sim}3$ larger than isotropic scattering and a factor of ${\sim}2$ larger than Rayleigh scattering. At the WFIRST/AFTA IWA of $0.2''$, this results in \fobs\,$\sim\!3\%$ for Rayleigh and isotropic scattering and \fobs\,$\sim\!10\%$ for Lambert scattering. Thus, when the geometric albedo is calculated self-consistently with the single-scattering albedo, the assumed scattering phase function can greatly influence the predicted yield from an optical direct-imaging survey. In particular, assuming the Lambert phase function generally results in an overestimate of the detectability of giant planets with respect to the more physically motivated Rayleigh phase function.

\subsubsection{Dependence on Planet Radius}

The planet/star flux ratio is proportional to the square of the planet radius. Therefore, we expect \fobs\ to be sensitive to small variations in planet size, which we have heretofore assumed to be that of Jupiter. Figure~\ref{fig:vary_Rp} 
shows \fobs\ as a function of planet radius, assuming observations made with a coronagraph that has $\m{IWA}=0.1'',\ 0.2'',\ \m{and}\ 0.3''$. All assumptions other than planet radius are as listed at the end of \textsection\ref{sec:mean_fobs}. We note that gas giants are expected to have $0.8~R_\m{Jup} \lesssim R_p \lesssim 1.2~R_\mathrm{Jup}$, but we show a wider range of radii for illustrative purposes. Within the range of expected radii for gas giants, an increase in radius by a factor of ${\sim}1.5$ leads to an increase in \fobs\ by a factor of ${\sim}1.5-2.5$, depending on the IWA. We find that smaller IWAs are less sensitive to variations in planet radius. This effect likely arises from the $d^{-2}$ dependence of the planet/star flux ratio, which---because of the increasing number of planets at smaller orbital distances that are accessible to smaller IWAs---reduces the weight carried by the $R_p^2$ dependence of the flux ratio.

\begin{figure}[t!]
\centering
\includegraphics[width=9.2cm, trim=0.3cm 0.cm -0.1cm 0cm,clip=true]{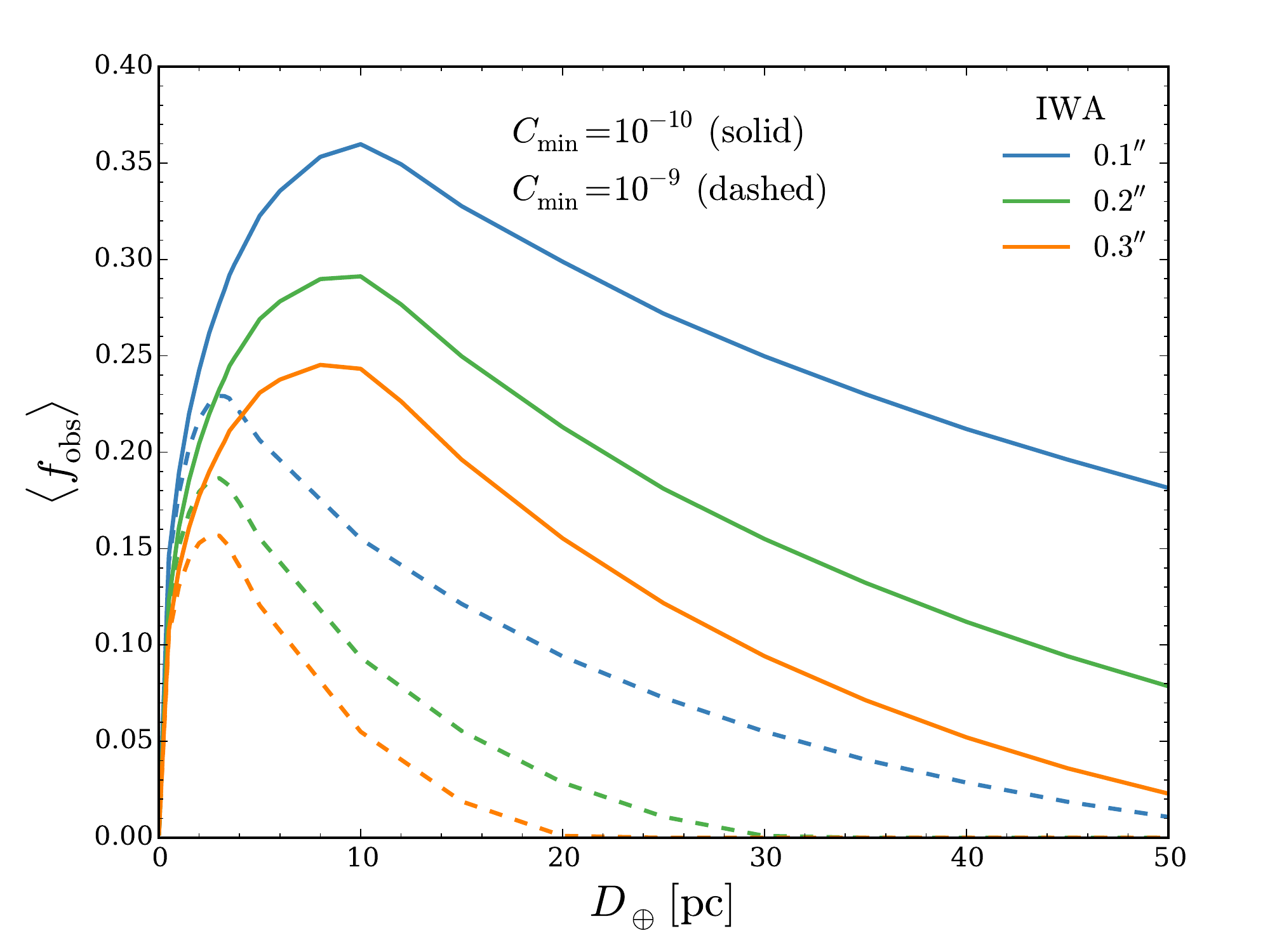}
\caption{The mean observability fraction \fobs\ as a function of distance from Earth ($D_\oplus$) and minimum achievable contrast ($C_\m{min}$), assuming the $\m{IWA}=0.1''$, $0.2''$, and $0.3''$. Dashed lines assume $C_\m{min}=10^{-9}$, and solid lines assume $C_\m{min}=10^{-10}$ . All other assumptions are as listed at the end of \textsection\ref{sec:mean_fobs}.
\vspace{0.08cm}}
\label{fig:vary_D_Cmin}
\end{figure}

\subsubsection{Dependence on Distance and Minimum Contrast}
The fiducial model listed at the end of \textsection\ref{sec:mean_fobs} assumes direct observations of planet/star systems located a distance $D_\oplus = 10$~pc from Earth with a minimum achievable contrast $C_\m{min} = 10^{-9}$. It is obvious that the direct detectability of EGPs will vary strongly with both of these parameters. In Figure~\ref{fig:vary_D_Cmin}, we investigate this variation in the context of the mean observability fraction. We show \fobs\ for planet/star systems located $0.01-50$~pc from Earth, assuming minimum achievable contrasts of $10^{-9}$ (dashed lines) and $10^{-10}$ (solid lines). The calculations assume a coronagraphic instrument with an $\m{IWA}=0.1'',\ 0.2'',\ \m{and}\ 0.3''$.  All other parameters are as listed at the end of \textsection\ref{sec:mean_fobs}. 

WFIRST/AFTA is anticipated to have $\m{IWA}\!\sim\!0.2''$ and $C_\m{min}\!\sim\!10^{-9}$. In this case, Jupiter-like planets located 10~pc from Earth spend on average ${\sim}10\%$ of their time in observable configurations, whereas the same planets located 30~pc from Earth spend on average ${\sim}0.1\%$ of their time in observable configurations. If instead a minimum contrast of $10^{-10}$ can be achieved, then these fractions become ${\sim}30\%$ and ${\sim}16\%$, respectively. At small distances from Earth, we see \fobs\ increases with distance. This trend is the result of planets at sufficiently wide orbital distances to spend some fraction of their orbit outside the OWA, which we have assumed to be $1.75''$. Assuming a uniform distribution of stars and $\m{IWA=0.2''}$, Jupiter-like planets orbiting stars within 10, 30, and 50~pc from Earth have volume-averaged observability fractions of ${\sim}12\%,\ 3\%,\ \m{and}\ 0.5\%$ for $C_\m{min}\!\sim\!10^{-9}$ and ${\sim}28\%,\ 20\%,\ \m{and}\ 13\%$ for $C_\m{min}\!\sim\!10^{-10}$.

\subsubsection{Dependence on the Semi-Major Axis Distribution}\label{sec:semi}
Our fiducial model assumes the semi-major axis distribution derived from radial-velocity planets can be extrapolated to wider separations ($a > 3$~AU). However, the validity of this assumption is highly uncertain, and, as evidenced by Figure~\ref{fig:mc_params}, the semi-major axis distribution has a large influence on the distribution of observability fractions. In this section, we explore the dependence of \fobs\ upon the semi-major axis distribution by varying each of its components: the minimum semi-major axis ($a_\m{min}$), the maximum semi-major axis ($a_\m{max}$), and the power-law index of the distribution~($\beta$).

\begin{figure}[b!]
\centering
\includegraphics[width=9.2cm, trim=0.3cm 0.cm 0.cm 0cm,clip=true]{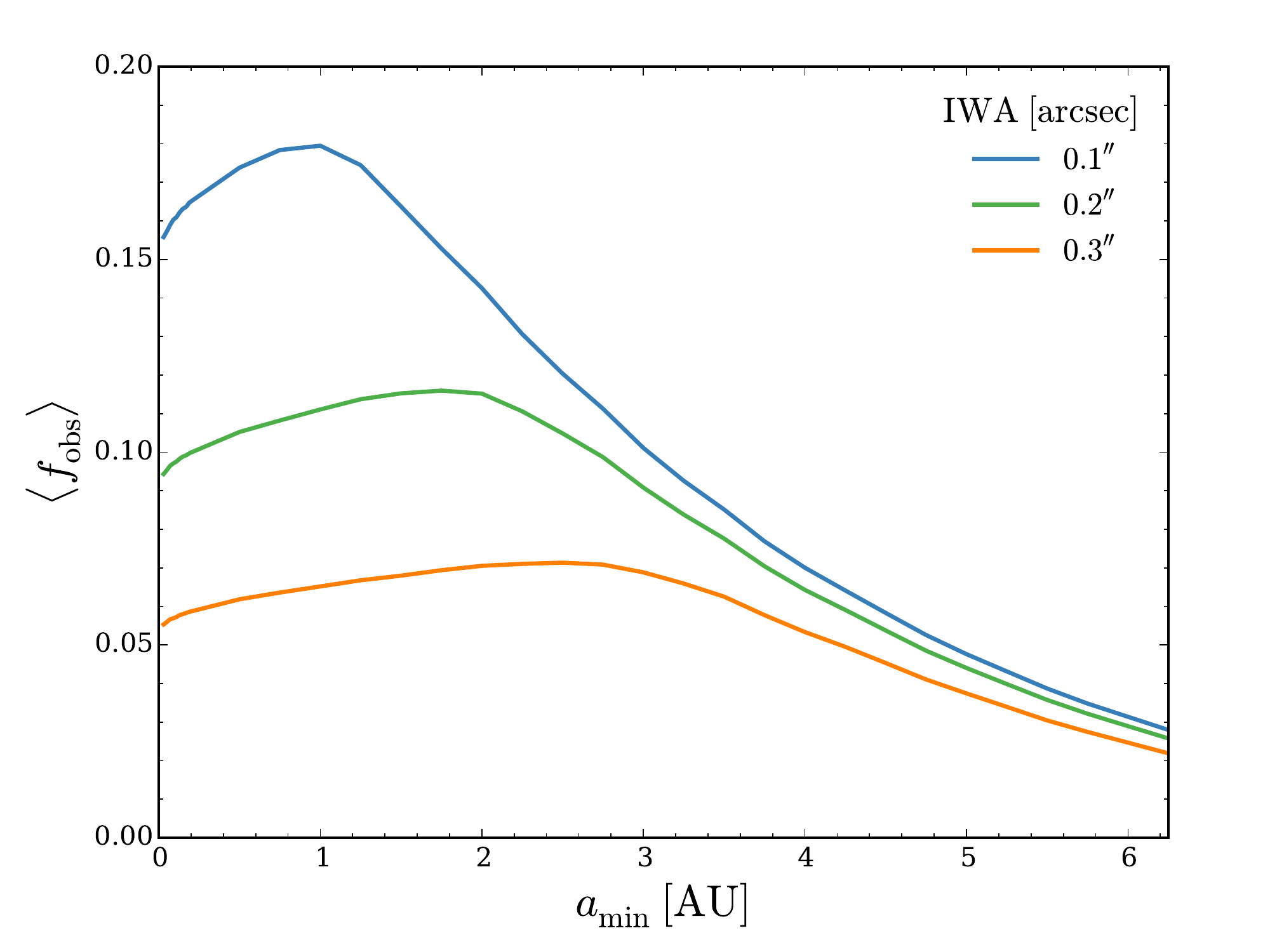}
\caption{Dependence of the mean observability fraction \fobs\ upon the minimum semi-major axis ($a_\m{min}$) of the semi-major axis distribution. All other model parameters are as listed at the end of \textsection\ref{sec:mean_fobs}.}
\label{fig:vary_amin}
\end{figure}

Our fiducial model assumes $a_\m{min}=0.03$~AU (note we set $e=0$ for any planet with $a<0.1$~AU), which has been inferred for radial-velocity planets \citep{Cumming2008}. In Figure~\ref{fig:vary_amin}, we show \fobs\ as a function of $a_\m{min}$, assuming the indicated IWAs. At a given IWA, the mean observability increases with increasing $a_\m{min}$ until $a_\m{min}/D_\oplus \sim \m{IWA}$, where $D_\oplus = 10$~pc in our fiducial model. Since all of the MC experiments use the same number of planets ($5\times10^5$), increasing the minimum semi-major axis forces more planets to exist at larger orbital separations, leading to larger \fobs\ when $a_\m{min}/D_\oplus \lesssim \m{IWA}$ and smaller \fobs\ when $a_\m{min}/D_\oplus \gtrsim \m{IWA}$. At the anticipated IWA of WFIRST/AFTA, the mean observability varies by ${\sim}1\%$ for $a_\m{min}$ in the range $0.03-2.0$~AU. 

\begin{figure}[t!]
\centering
\includegraphics[width=9.2cm, trim=0.3cm 0.cm 0.cm 0cm,clip=true]{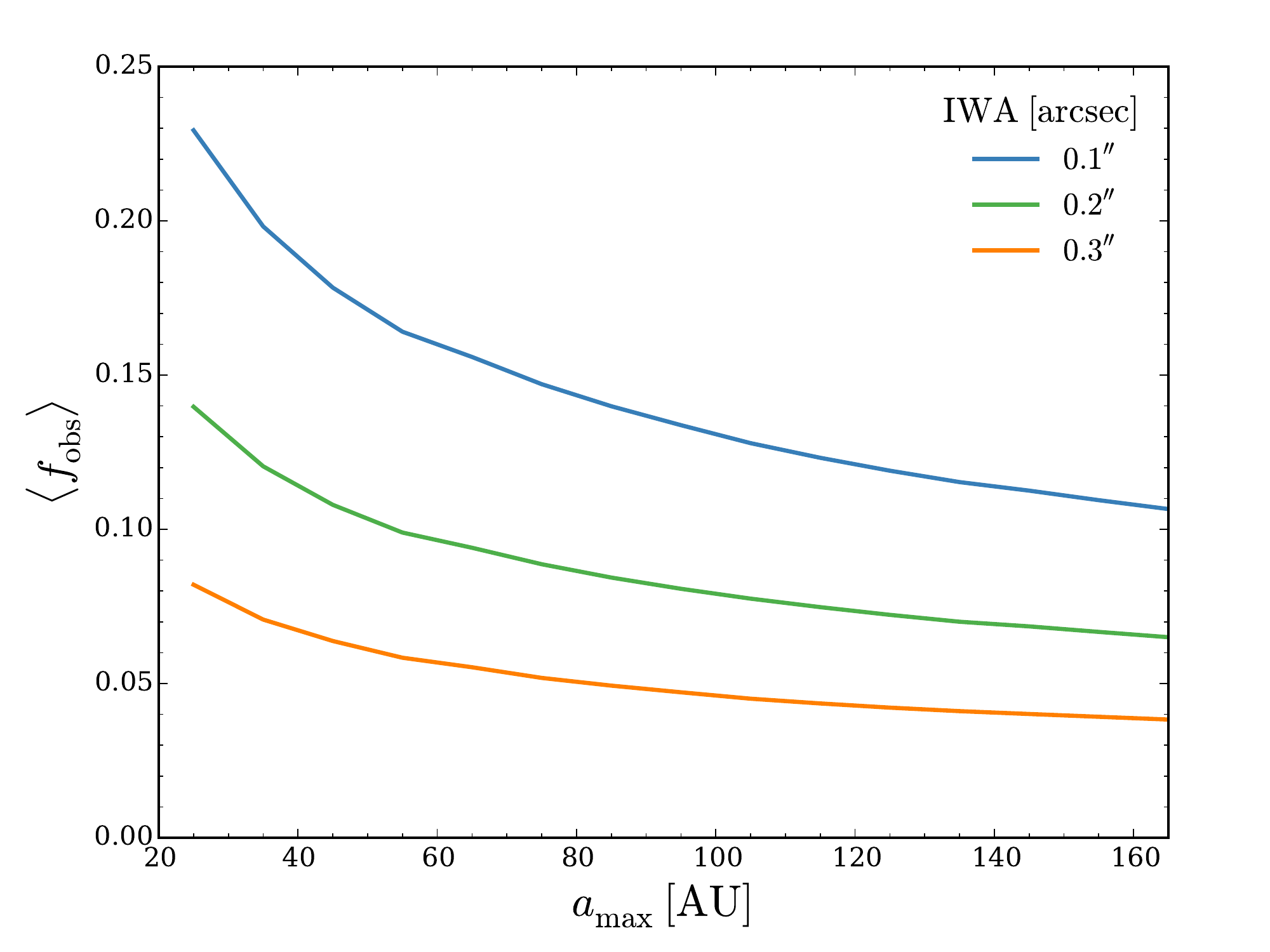}
\caption{Dependence of the mean observability fraction \fobs\ upon the maximum semi-major axis ($a_\m{max}$) of the semi-major axis distribution. All other model parameters are as listed at the end of \textsection\ref{sec:mean_fobs}.}
\label{fig:vary_amax}
\end{figure}
\begin{figure}[t!]
\centering
\includegraphics[width=9.2cm, trim=0.3cm 0.cm -0.3cm 0cm,clip=true]{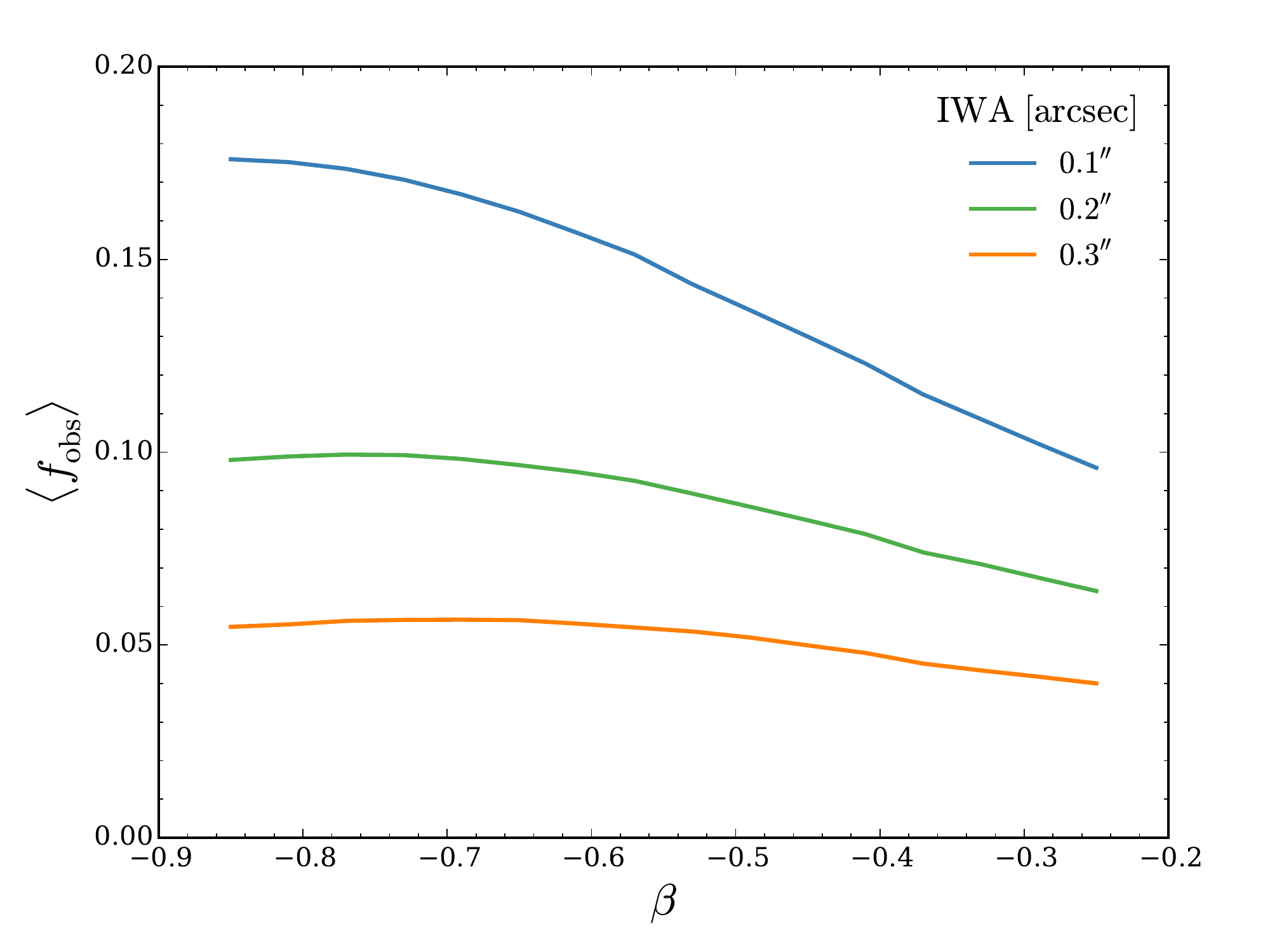}
\caption{Dependence of the mean observability fraction \fobs\ upon the power-law index of the semi-major axis distribution ($\beta$). All other model assumptions are as listed at the end of \textsection\ref{sec:mean_fobs}.}
\label{fig:vary_beta}
\end{figure}

As mentioned in \textsection\ref{sec:param_dist}, recent model-dependent limits on $a_\m{max}$ fall in the range ${\sim}30-100$~AU \citep{Brandt2014}. In Figure~\ref{fig:vary_amax}, we test the sensitivity of \fobs\ to a somewhat larger range of $a_\m{max}$. For a fixed number of planets, smaller $a_\m{max}$ values yield larger \fobs\ values, which is sensible given that more planets will exist at smaller, but still detectable, orbital distances. As $a_\m{max}$ increases, \fobs\ slowly decreases as a result of the increasing number of planets at very large orbital distances, which are too faint to be detected and, in the case of nearly face-on orbits, fall outside the OWA. Over the range $a_\m{max}\sim30-100$~AU, \fobs\ decreases by a factor of ${\sim}1.6$ for each IWA. 

The power-law index of the semi-major axis distribution of radial-velocity planets with periods in the range $2-2000$~days has been constrained to be $\beta=-0.61\pm0.15$ \citep{Cumming2008}. Our fiducial model assumes this index can be extrapolated to planets at much larger orbital separations. In Figure~\ref{fig:vary_beta}, we test the sensitivity of \fobs\ to variations in the power-law index. As we steepen $\beta$ towards more negative values, an increasing number of planets begin to pile up at small orbital separations, which leads to larger \fobs\ values. At the anticipated IWA of WFIRST/AFTA, $\beta=-0.85$ and $\beta=-0.61$, which are the estimated indices from \citet{Brandt2014} and \citet{Cumming2008}, produce \fobs\ values that are nearly identical. Over the full range of $\beta$, the difference between the minimum and maximum \fobs\ values is only ${\sim}3\%$. 

\subsubsection{Geometric Albedo as a Function of Distance} \label{sec:Agdist}

Thus far, we have assumed the geometric albedo is constant throughout the orbit of each planet. However, the atmospheric temperature of a planet varies strongly with orbital distance, and such variations determine whether water or ammonia clouds will form. We, therefore, expect the albedo spectrum to vary significantly with orbital distance \citep{Sudarsky2005}. We also expect the albedo spectrum to depend sensitively upon the planet's mass, age, and parent star. In this section, we compare our fiducial assumption of $A_g = 0.5 =$~constant with two simple models that account for the distance dependence of the albedo spectrum. 

\begin{figure}[t!]
\centering
\includegraphics[width=9cm, trim=0.7cm 0cm 0.6cm 0cm,clip=true]{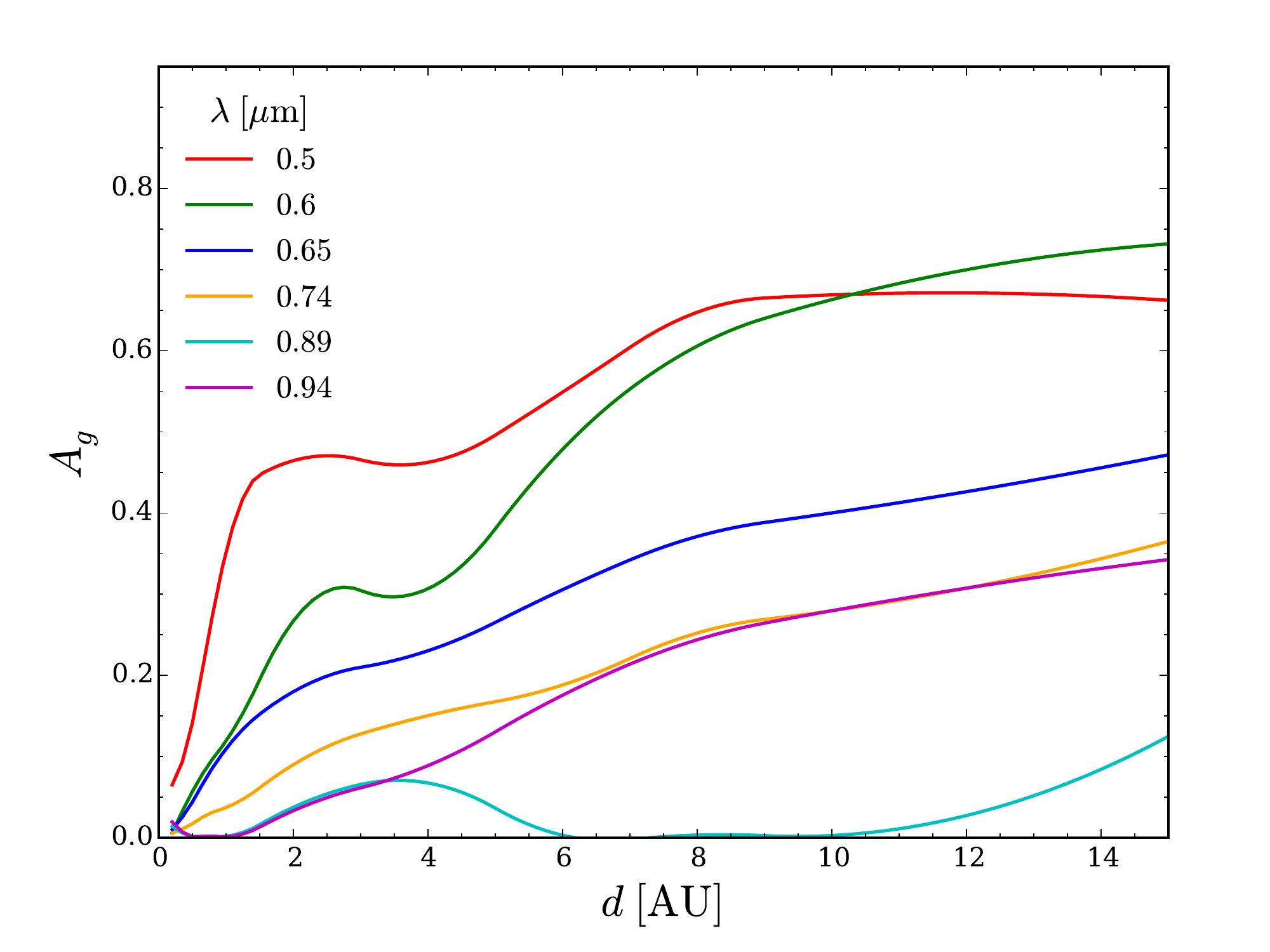}
\caption{The geometric albedo as a function of distance and wavelength for a Jupiter-mass, 5-Gyr planet orbiting a G2~V star. These results are an interpolation of the models shown in Figure~9 of \citet{Sudarsky2005}.}
\label{fig:Ag_dist}
\end{figure}

The first model is an interpolation of the models shown in Figure~9 of \citet{Sudarsky2005}, which depicts the geometric albedo spectra of a Jupiter-mass, 5-Gyr planet orbiting a G2~V star at distances in the range $0.2-15$~AU. In Figure~\ref{fig:Ag_dist}, we show the predictions of this model for the distance dependence of the geometric albedo for various wavelengths, including wavelengths associated with methane ($0.74~\mu$m and $0.89~\mu$m), ammonia ($0.65~\mu$m), and water ($0.94~\mu$m). We see the geometric albedo, and hence the planet/star flux ratio, is a strong, non-monotonic function of both distance and wavelength. 

\begin{figure}[t!]
\centering
\includegraphics[width=9.2cm, trim=0.58cm 0.cm 0.cm 0cm,clip=true]{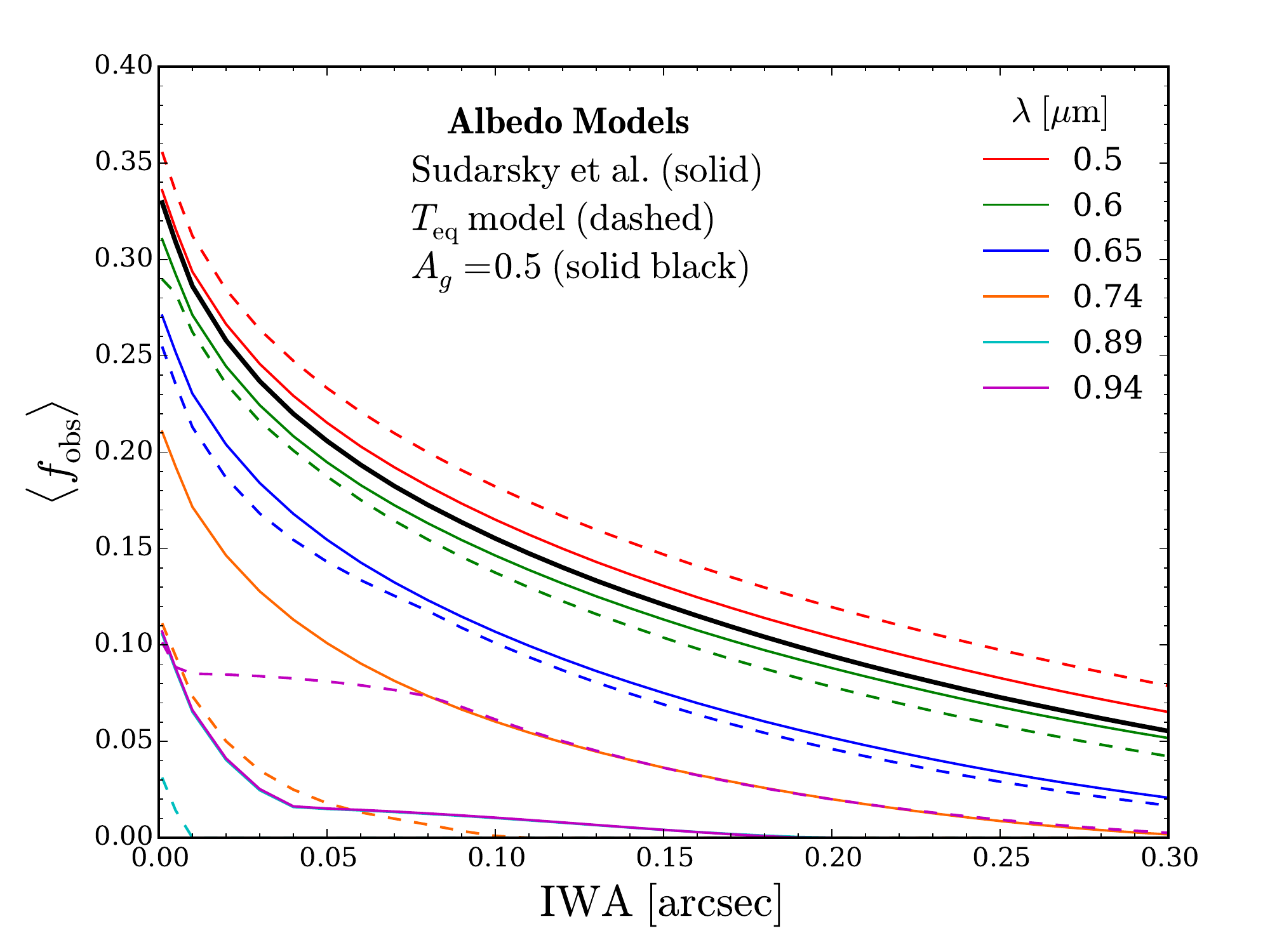}
\caption{Comparison of \fobs\ assuming three models for the albedo spectrum. The ``Sudarsky et al.'' (solid lines) and ``$T_\m{eq}$'' (dashed lines) models account for the distance dependence of the albedo spectrum, whereas the $A_g=0.5$ model (solid black line), which is our fiducial model, holds it constant throughout each planet's orbit. The solid cyan and purple lines are overlapping, which indicates that orbital distances $d<4$~AU are dominating the mean observability at these wavelengths (see Figure~\ref{fig:Ag_dist}).}
\label{fig:Agmodels}
\end{figure}

Using this model for $A_g$, we carried out MC experiments for observations at the wavelengths shown in Figure~\ref{fig:Ag_dist}, keeping all other parameters as listed at the end of \textsection\ref{sec:mean_fobs}. For orbital distances $d>15$~AU, we set $A_g$ equal to its value at $15$~AU; this does not significantly influence our results, since planets at these separations contribute very little to \fobs\ (see Figure~\ref{fig:mc_params}). In Figure~\ref{fig:Agmodels}, we show \fobs\ as a function of IWA assuming the ``Sudarsky et al.'' albedo model (solid lines) and our fiducial assumption of $A_g = 0.5 = $ constant (solid black line). As can be seen in the first column of Figure~\ref{fig:mc_params}, most planets with nonzero $f_\m{obs}$ values have semi-major axes in the range $2-10$~AU, so it is the variability of $A_g$ in this region that has the strongest effect on \fobs. In the wavelength range we have considered, this model suggests the optimal observability fractions occur for $\lambda \sim 0.5-0.6~\m{\mu m}$, which Figure~\ref{fig:Agmodels} indicates is the effective bandpass of our fiducial assumption $A_g=0.5$. At the IWA of WFIRST/AFTA, direct observations in the wavelength range $\lambda \sim 0.5-0.6~\m{\mu m}$ of a Jupiter-like planet located 10~pc from Earth will, on average, be possible for ${\sim}10\%$ of its orbital period. We note that the observed reflected fluxes shortward of ${\sim}0.6~\m{\mu m}$ from Jupiter and Saturn are less than what this model would predict, which is made manifest by the reddish hues of these solar-system planets \citep{Karkoschka1994, Burrows2014}. 


For the second model of the distance dependence of the albedo spectrum, we calculate the equilibrium temperature for each planet as a function of orbital distance in a self-consistent fashion with the albedo spectra. This is accomplished as follows. For each planet in an MC experiment, we assume the planetary atmosphere is homogeneous and Rayleigh scattering. In addition, solar metallicity and a pressure of $0.5$~bar are assumed (albedo spectra for such a planet with $T=200$~K are shown as black lines in the left column of Figure~\ref{fig:scat_albedo}). At a given phase of the planet's orbit, we assume an initial Bond albedo and calculate the equilibrium temperature. Assuming this to be the temperature of the atmosphere, we generate (scattering, geometric, and spherical) albedo spectra with the methods described in \textsection\ref{sec:atms}. We then integrate the spherical albedo spectrum to obtain the Bond albedo and calculate a new equilibrium temperature ($T_\m{eq}$). We iterate this process until we have achieved self-consistency between the equilibrium temperature and albedo spectra. This procedure is performed at every phase of the planet's orbit, resulting in a geometric albedo spectrum and phase function that varies with orbital distance. The results of MC experiments assuming this ``$T_\m{eq}$ model'' are shown as dashed-lines in Figure~\ref{fig:Agmodels}. We see that this model is roughly consistent with the Sudarsky et al. model for $\lambda < 0.7~ \m{\mu m}$, but its predictions are very different at longer wavelengths. This model also predicts that the wavelength range $\lambda \sim 0.5-0.6~\m{\mu m}$ offers the optimal observability fractions. 

\subsection{Detection Probabilities}\label{sec:probs}

Given assumptions about the orbital parameter distributions and occurrence rates ($f_p$) of a population of planets, the mean observability fractions presented in this work can be translated into the probability that a blind search for such planets will yield a detection. Statistical studies of planets discovered by the radial-velocity and microlensing methods suggest that the frequency of giant planets with $a < 20$~AU falls in the range $f_p \sim 10 - 20\%$ \citep{Cumming2008, Cassan2012}. If we assume $f_p=15\%$, then calculations based on our fiducial model assumptions, which are listed at the end of \textsection\ref{sec:mean_fobs}, suggest that the probability of detecting a Jupiter-like planet in a blind search around a star located 10~pc from Earth is $p = f_p \cdot \langle f_{obs} \rangle \sim 1.5\%$, where we have assumed $\m{IWA}=0.2''$. If a minimum contrast of $10^{-10}$ is assumed, then this probability increases to $p\sim4.2\%$. If the star is located 20~pc from Earth, we find $p\sim0.4\%$ for $C_\m{min}=10^{-9}$ and $p=3.2\%$ for $C_\m{min}=10^{-10}$. The scaling of $p$ with parameters such as the IWA, planet radius, distance from Earth, and atmospheric scattering properties can be inferred from the figures presented in this paper. In all but the most optimistic configurations, the probability for detection in a blind search carried out with WFIRST/AFTA's baseline coronagraphic capabilities is low ($<5\%$).

By the time direct detection in the optical becomes feasible, long-term radial-velocity campaigns will have likely discovered many wide-separation EGPs, for which a subset of Keplerian elements will be known. Further, it has been estimated that {\it Gaia} will discover and determine robust orbits for ${\sim}1500$ planets with periods $\lesssim6$~yr within 50~pc from Earth \citep{Perryman2014}, including ${\sim}100$ giant planets around known M dwarf host stars within 30~pc \citep{Sozzetti2014}. Direct-imaging surveys that leverage this information can improve their detection probabilities with respect to blind searches, and the tools we have developed for this work can be used to determine both the most promising systems to target and when to observe them. It is important to point out that,  as can be inferred from Figures~$\ref{fig:i-e_fr}-\ref{fig:observe_frac}$, the window of time that an EGP is directly detectable can be very brief and sparsely distributed throughout its orbit, suggesting that previous orbital parameter constraints must be very precise if they are to be used to optimize direct-imaging surveys. For a detailed study of the impact of uncertain radial-velocity orbital parameters on direct imaging and spectroscopy of radial-velocity exoplanets, see \citet{Brown2015}.

\begin{figure}[t!]
\centering
\includegraphics[width=9.2cm, trim=0.97cm 0.cm 0.1cm 0cm,clip=true]{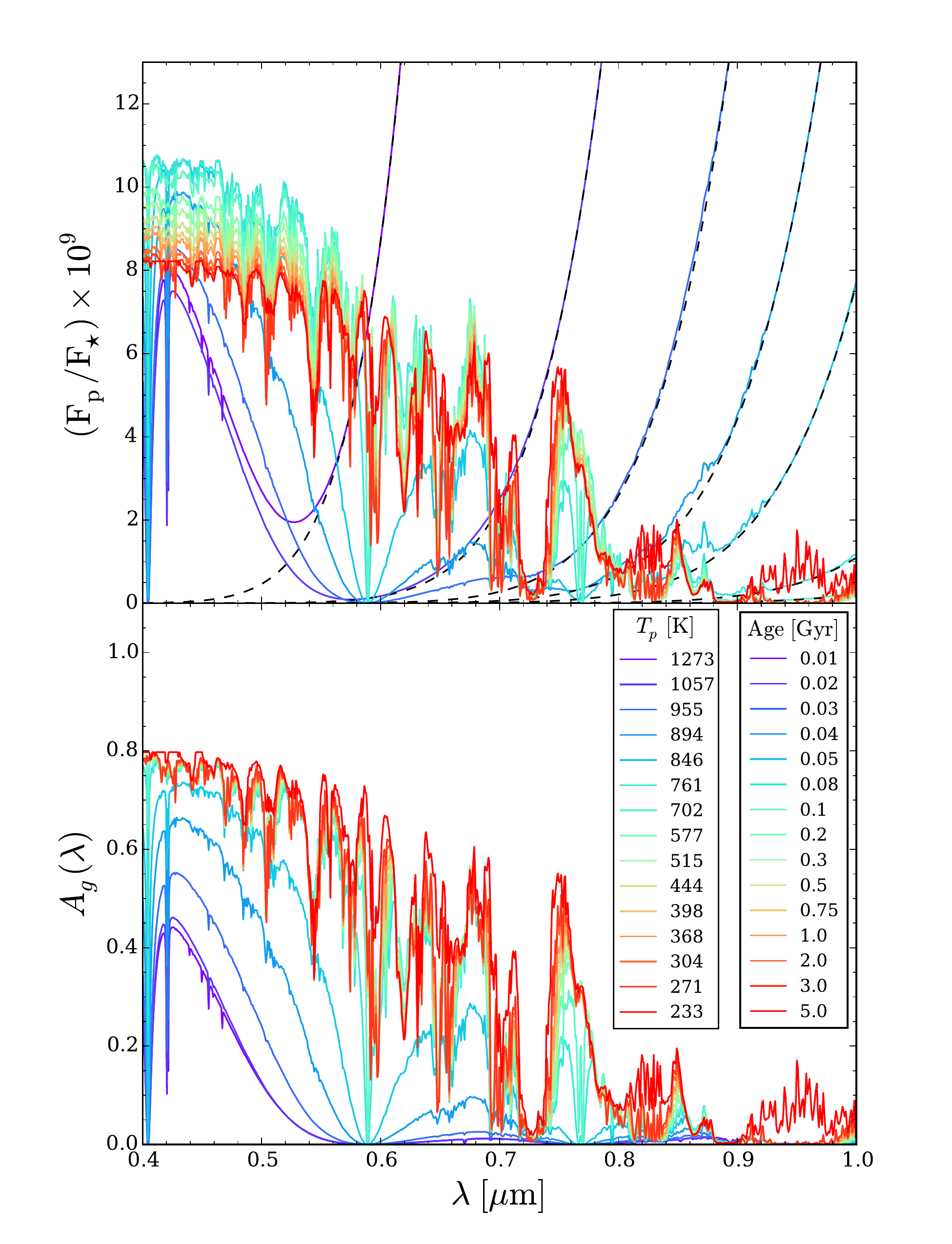}
\caption{The planet/star flux ratio (top) and geometric albedo spectrum (bottom), as a function of wavelength and age, for a 7 Jupiter-mass planet orbiting a Sun-like star at an orbital distance of 5 AU, where we have included the effect of blackbody thermal emission from the planet. The black dashed lines in the top panel show the component of the contrast ratio that is associated with thermal emission from the planet. We use the evolutionary calculations of \citet{Burrows1997} to model the evolution of the age, radius, and temperature of the planet. The atmospheric pressure is 0.5~bar, and solar metallically is assumed. We have assumed both the star and planet are uniformly luminous blackbodies. The albedo spectrum was calculated with the opacity database from \citet{SharpBurrows2007} and the equilibrium chemical abundances from \citet{BurrowsSharp1999}. The atmosphere is assumed to be Rayleigh scattering. The planet age and effective temperature ($T_p$) color-scales  apply to both panels.}
\label{fig:fr_age}
\end{figure}

\subsection{Thermal Emission from Young Planets}\label{sec:thermal}
Throughout this work, we have focused on cool, relatively old giant planets, for which the flux from thermal emission is negligible compared to the flux from reflected starlight. The thermal emission from a giant planet is age, mass, and metallicity dependent. Furthermore, the particular formation mechanism of giant planets significantly influences the cooling of the planet \citep[e.g.,][]{Burrows1997, Marley2007, Fortney2008, Spiegel2012}. Regardless of the assumed cooling model, it is reasonable to expect the thermal emission contribution to the optical planet/star flux ratio will increase with planet mass and decrease with planet age. 

In Figure~\ref{fig:fr_age}, we show the planet/star flux ratio (top panel) and associated Rayleigh scattering geometric albedo spectrum (bottom panel) for a representative 7 Jupiter-mass planet orbiting a Sun-like star at an orbital distance of 5~AU, where we have included the effect of blackbody thermal emission from the planet. We use the evolutionary calculations of \citet{Burrows1997} to model the evolution of the age, radius, and temperature of the planet, and the albedo spectrum was calculated with the opacity database from \citet{SharpBurrows2007} and  equilibrium chemical abundances from \citet{BurrowsSharp1999}. We assume solar metallically and an atmospheric pressure of 0.5~bar. Both the planet and star are assumed to be uniformly luminous blackbodies. That is, we assume the thermal emission component of the contrast ratio is given by $R^2_p B_\lambda(T_p) / R^2_\star B_\lambda(T_\star)$, where $T_p$ is the effective temperature of the planet and $B_\lambda(T)$ is the Planck function. 

In this simple model, the thermal emission from the planet noticeably contributes to the optical planet/star flux ratio only for ages $<80$~Myr. Nonetheless, we see it briefly dominates the flux ratio at wavelengths $\gtrsim0.5~\m{\mu m}$ when the planet is a mere few tens of millions of years old, resulting in a planet that transitions from red to blue with increasing age and decreasing effective temperature. The age range over which thermal emission is important will be larger (smaller) for giant planets more (less) massive than 7~Jupiter masses. The decrease in flux at short wavelengths (${\sim}0.4-0.5~\m{\mu m}$) for ages $\gtrsim100$~Myr is due primarily to the contraction of the planet's radius, which decreases from ${\sim}1.4~R_\m{Jup}$ to ${\sim}1~R_\m{Jup}$ throughout its evolution. In contrast, the geometric albedo spectrum is approximately constant over this same wavelength and age interval. 

\section{Summary and Conclusions}\label{sec:summary}

With the planning of future exoplanet-focused missions such as WFIRST/AFTA well underway, it seems likely that the direct detection of EGPs in the optical will become feasible in the coming decade. Therefore, studies of the observational signatures and direct detectability of reflected light from EGPs are particularly timely and will aid in the design and optimization of these missions. In this work, we have presented an exploration of the direct detectability of EGPs at visible wavelengths, quantified by the fraction of a planet's orbit for which it is in an observable configuration---its observability fraction $f_\m{obs}$. Using a suite of Monte Carlo experiments, we investigated the dependence of $f_\m{obs}$ upon various technological and astrophysical parameters such as the inner working angle and minimum achievable contrast of the direct-imaging observatory; the planet's scattering phase function, geometric albedo, single-scattering albedo, radius, and distance from Earth; and the semi-major axis distribution of EGPs. Unless stated otherwise, we now summarize our results assuming observations of Jupiter-like planets\footnote{We remind the reader that we define a ``Jupiter-like'' planet to be an EGP with the radius of Jupiter, $A_g=0.5$, and a Rayleigh scattering atmosphere.}, using an optical coronagraphic instrument with $\m{IWA}=0.2''$ and $C_\m{min}=10^{-9}$, which are the baseline design parameters for WFIRST/AFTA.

Low-inclination orbits are highly favorable for direct-imaging observations (Figures~\ref{fig:i-e_fr} and \ref{fig:mc_params}). However, planets with $i\gtrsim50^\circ$ and $\omega_p\sim270^\circ$ (Figures~\ref{fig:i-wp_fr} and \ref{fig:mc_params}) are generally observable out to larger semi-major axes than planets in other configurations. At a distance of 10~pc from Earth, only (low- to modest-eccentricity) EGPs with semi-major axes in the range $a\sim2 - 10$~AU will have $f_\m{obs}>1\%$. If $C_\m{min}\sim10^{-10}$ can be achieved, this range extends out to $a\sim35$~AU, with a much larger fraction of orbital parameter space yielding $f_\m{obs}\sim1$  (Figure~\ref{fig:mc_params}). Inspection of the bottom row of  Figure~\ref{fig:mc_params} suggests that the mean direct detectability of EGPs is relatively insensitive to their eccentricity distribution. 

The mean observability fraction, \fobs, of EGPs is a strong function of their atmospheric scattering properties. For a given geometric albedo, Rayleigh scattering atmospheres predict \fobs\ values that are generally ${\sim}20\%$ below the predictions of isotropic and Lambert scattering atmospheres (left panel, Figure~\ref{fig:fix_Ag_om}). If instead one assumes a single-scattering albedo and self-consistently calculates the phase function for a given scattering mechanism, the often-assumed Lambertian phase function can overestimate the detectability of EGPs by as much as a factor of ${\sim}4$ with respect to the more physically motivated Rayleigh phase function (right panel, Figure~\ref{fig:fix_Ag_om}).

Within the range of expected radii for gas giants ($0.8\,R_\m{Jup} \lesssim R_p \lesssim 1.2\,R_\m{Jup}$), an increase in radius by a factor of ${\sim}1.5$ leads to an increase in \fobs\ by a factor of ${\sim}1.8$ (Figure~\ref{fig:vary_Rp}). We find that smaller IWAs are less sensitive to variations in planet radius. This effect likely arises from the $d^{-2}$ dependence of the planet/star flux ratio, which---because of the increasing number of planets at smaller separations that are accessible to smaller IWAs---reduces the weight carried by the $R_p^2$ dependence of the flux ratio.

An EGP located 10~pc from Earth will spend on average ${\sim}10\%$ of its time in observable configurations, whereas the same planet located 30~pc from Earth will spend on average ${\sim}0.1\%$ of its time in observable configurations. If a minimum contrast of $10^{-10}$ can be achieved, then these fractions become ${\sim}30\%$ and ${\sim}16\%$, respectively (Figure~\ref{fig:vary_D_Cmin}). Assuming a uniform distribution of stars, EGPs orbiting stars within 10, 30, and 50~pc from Earth have volume-averaged observability fractions of ${\sim}12\%,\ 3\%,\ \m{and}\ 0.5\%$ for $C_\m{min}\!\sim\!10^{-9}$ and ${\sim}28\%,\ 20\%,\ \m{and}\ 13\%$ for $C_\m{min}\!\sim\!10^{-10}$.  Thus, the scientific return from a future optical direct-imaging survey will depend critically upon its minimum achievable contrast, as well as the number of wide-separation EGPs that exist within a few tens of parsecs from Earth. 

In \textsection\ref{sec:semi} we explored the dependence of \fobs\ upon the semi-major axis distribution of EGPs by varying each of its components: the minimum semi-major axis ($a_\m{min}$), the maximum semi-major axis ($a_\m{max}$), and the power-law index of the distribution~($\beta$). Given constraints from radial-velocity planets and infrared direct-imaging surveys, the uncertainty in \fobs\ is dominated by the uncertainty in $a_\m{max}$. For $a_\m{max}$ in the range $30-100$~AU, which is the model-dependent limit set by \citet{Brandt2014}, \fobs\ varies by a factor of ${\sim}1.6$ (Figure~\ref{fig:vary_amax}).

The albedo spectrum of a giant planet is expected to be a non-monotonic function of both distance and wavelength \citep[Figure~\ref{fig:Agmodels} and][]{Sudarsky2005}. Using two simple models for the distance- and wavelength-dependence of the geometric albedo, we find that bluer wavelengths result in larger \fobs\ values, which is to be expected given that we are interested in reflected-light observations. We caution that, as revealed by their reddish hues, the observed reflected fluxes shortward of ${\sim}0.6~\m{\mu m}$ from Jupiter and Saturn are less than what our models would predict. 

As described in \textsection\ref{sec:probs}, the mean observability fraction for a population of planets is proportional to the probability ($p$) that a blind search for such planets will yield a detection: $p = f_p \cdot \langle f_{obs} \rangle$, where $f_p$ is the planet occurrence rate. The scaling of $p$ with parameters such as IWA, $C_\m{min}$, planet radius, distance from Earth, and atmospheric scattering properties can be inferred from the figures presented in this paper. In all but the most optimistic configurations, the probability for detection in a blind search carried out with WFIRST/AFTA's baseline coronagraphic capabilities is low ($<5\%$).

Finally, we note that thermal emission from young EGPs can provide a non-negligible contribution to optical planet/star flux ratios. Assuming the evolutionary calculations of \citet{Burrows1997},  the thermal emission from a 7 Jupiter-mass planet orbiting a Sun-like star briefly dominates the flux ratio at wavelengths $\gtrsim0.5~\m{\mu m}$ when the planet is a few tens of millions of years old, becoming completely negligible after ${\sim}80$~Myr. The result is a planet that transitions from red to blue with increasing age and decreasing effective temperature (Figure~\ref{fig:fr_age}). The age range over which the thermal emission contribution to the optical flux ratio is significant will be larger (smaller) for giant planets more (less) massive than the representative 7 Jupiter-mass planet we have considered here. 

\acknowledgements
JPG is supported by the National Science Foundation Graduate Research Fellowship  under Grant No. DGE 1148900. A.B. would like to acknowledge support in part under NASA grant NNX15AE19G, JPL subcontract no. 1513640, NASA HST awards HST-GO-13467.10-A, HST-GO-12550.02, and HST-GO-12473.06-A, and JPL/Spitzer contracts 1439064 and 1477493.  
\bibliographystyle{apj}
\bibliography{mybib}

\begin{thebibliography}{}
\expandafter\ifx\csname natexlab\endcsname\relax\def\natexlab#1{#1}\fi

\bibitem[{{Barman} {et~al.}(2011){Barman}, {Macintosh}, {Konopacky}, \&
  {Marois}}]{Barman2011}
{Barman}, T.~S., {Macintosh}, B., {Konopacky}, Q.~M., \& {Marois}, C. 2011,
  \apj, 733, 65

\bibitem[{{Batalha} {et~al.}(2013){Batalha}, {Rowe}, {Bryson}, {Barclay},
  {Burke}, {Caldwell}, {Christiansen}, {Mullally}, {Thompson}, {Brown},
  {Dupree}, {Fabrycky}, {Ford}, {Fortney}, {Gilliland}, {Isaacson}, {Latham},
  {Marcy}, {Quinn}, {Ragozzine}, {Shporer}, {Borucki}, {Ciardi}, {Gautier},
  {Haas}, {Jenkins}, {Koch}, {Lissauer}, {Rapin}, {Basri}, {Boss}, {Buchhave},
  {Carter}, {Charbonneau}, {Christensen-Dalsgaard}, {Clarke}, {Cochran},
  {Demory}, {Desert}, {Devore}, {Doyle}, {Esquerdo}, {Everett}, {Fressin},
  {Geary}, {Girouard}, {Gould}, {Hall}, {Holman}, {Howard}, {Howell},
  {Ibrahim}, {Kinemuchi}, {Kjeldsen}, {Klaus}, {Li}, {Lucas}, {Meibom},
  {Morris}, {Pr{\v s}a}, {Quintana}, {Sanderfer}, {Sasselov}, {Seader},
  {Smith}, {Steffen}, {Still}, {Stumpe}, {Tarter}, {Tenenbaum}, {Torres},
  {Twicken}, {Uddin}, {Van Cleve}, {Walkowicz}, \& {Welsh}}]{Batalha2013}
{Batalha}, N.~M., {Rowe}, J.~F., {Bryson}, S.~T., {et~al.} 2013, \apjs, 204, 24

\bibitem[{{Beichman} {et~al.}(2010){Beichman}, {Krist}, {Trauger}, {Greene},
  {Oppenheimer}, {Sivaramakrishnan}, {Doyon}, {Boccaletti}, {Barman}, \&
  {Rieke}}]{Beichman2010}
{Beichman}, C.~A., {Krist}, J., {Trauger}, J.~T., {et~al.} 2010, \pasp, 122,
  162

\bibitem[{{Beuzit} {et~al.}(2008){Beuzit}, {Feldt}, {Dohlen}, {Mouillet},
  {Puget}, {Wildi}, {Abe}, {Antichi}, {Baruffolo}, {Baudoz}, {Boccaletti},
  {Carbillet}, {Charton}, {Claudi}, {Downing}, {Fabron}, {Feautrier},
  {Fedrigo}, {Fusco}, {Gach}, {Gratton}, {Henning}, {Hubin}, {Joos}, {Kasper},
  {Langlois}, {Lenzen}, {Moutou}, {Pavlov}, {Petit}, {Pragt}, {Rabou}, {Rigal},
  {Roelfsema}, {Rousset}, {Saisse}, {Schmid}, {Stadler}, {Thalmann}, {Turatto},
  {Udry}, {Vakili}, \& {Waters}}]{Beuzit2008}
{Beuzit}, J.-L., {Feldt}, M., {Dohlen}, K., {et~al.} 2008, in Society of
  Photo-Optical Instrumentation Engineers (SPIE) Conference Series, Vol. 7014,
  Society of Photo-Optical Instrumentation Engineers (SPIE) Conference Series,
  18

\bibitem[{{Brandt} {et~al.}(2014){Brandt}, {McElwain}, {Turner}, {Mede},
  {Spiegel}, {Kuzuhara}, {Schlieder}, {Wisniewski}, {Abe}, {Biller},
  {Brandner}, {Carson}, {Currie}, {Egner}, {Feldt}, {Golota}, {Goto}, {Grady},
  {Guyon}, {Hashimoto}, {Hayano}, {Hayashi}, {Hayashi}, {Henning}, {Hodapp},
  {Inutsuka}, {Ishii}, {Iye}, {Janson}, {Kandori}, {Knapp}, {Kudo}, {Kusakabe},
  {Kwon}, {Matsuo}, {Miyama}, {Morino}, {Moro-Mart{\'{\i}}n}, {Nishimura},
  {Pyo}, {Serabyn}, {Suto}, {Suzuki}, {Takami}, {Takato}, {Terada}, {Thalmann},
  {Tomono}, {Watanabe}, {Yamada}, {Takami}, {Usuda}, \& {Tamura}}]{Brandt2014}
{Brandt}, T.~D., {McElwain}, M.~W., {Turner}, E.~L., {et~al.} 2014, \apj, 794,
  159

\bibitem[{{Brown}(2015)}]{Brown2015}
{Brown}, R.~A. 2015, \apj, 805, 188

\bibitem[{{Buenzli} \& {Schmid}(2009)}]{Buenzli2009}
{Buenzli}, E., \& {Schmid}, H.~M. 2009, \aap, 504, 259

\bibitem[{{Burrows}(2014)}]{Burrows2014}
{Burrows}, A. 2014, ArXiv e-prints, arXiv:1412.6097

\bibitem[{{Burrows} \& {Sharp}(1999)}]{BurrowsSharp1999}
{Burrows}, A., \& {Sharp}, C.~M. 1999, \apj, 512, 843

\bibitem[{{Burrows} {et~al.}(2004){Burrows}, {Sudarsky}, \&
  {Hubeny}}]{Burrows2004}
{Burrows}, A., {Sudarsky}, D., \& {Hubeny}, I. 2004, \apj, 609, 407

\bibitem[{{Burrows} {et~al.}(1997){Burrows}, {Marley}, {Hubbard}, {Lunine},
  {Guillot}, {Saumon}, {Freedman}, {Sudarsky}, \& {Sharp}}]{Burrows1997}
{Burrows}, A., {Marley}, M., {Hubbard}, W.~B., {et~al.} 1997, \apj, 491, 856

\bibitem[{{Cahoy} {et~al.}(2010){Cahoy}, {Marley}, \& {Fortney}}]{Cahoy2010}
{Cahoy}, K.~L., {Marley}, M.~S., \& {Fortney}, J.~J. 2010, \apj, 724, 189

\bibitem[{{Cassan} {et~al.}(2012){Cassan}, {Kubas}, {Beaulieu}, {Dominik},
  {Horne}, {Greenhill}, {Wambsganss}, {Menzies}, {Williams}, {J{\o}rgensen},
  {Udalski}, {Bennett}, {Albrow}, {Batista}, {Brillant}, {Caldwell}, {Cole},
  {Coutures}, {Cook}, {Dieters}, {Prester}, {Donatowicz}, {Fouqu{\'e}}, {Hill},
  {Kains}, {Kane}, {Marquette}, {Martin}, {Pollard}, {Sahu}, {Vinter},
  {Warren}, {Watson}, {Zub}, {Sumi}, {Szyma{\'n}ski}, {Kubiak}, {Poleski},
  {Soszynski}, {Ulaczyk}, {Pietrzy{\'n}ski}, \& {Wyrzykowski}}]{Cassan2012}
{Cassan}, A., {Kubas}, D., {Beaulieu}, J.-P., {et~al.} 2012, \nat, 481, 167

\bibitem[{{Chauvin} {et~al.}(2004){Chauvin}, {Lagrange}, {Dumas}, {Zuckerman},
  {Mouillet}, {Song}, {Beuzit}, \& {Lowrance}}]{Chauvin2004}
{Chauvin}, G., {Lagrange}, A.-M., {Dumas}, C., {et~al.} 2004, \aap, 425, L29

\bibitem[{{Chauvin} {et~al.}(2005){Chauvin}, {Lagrange}, {Zuckerman}, {Dumas},
  {Mouillet}, {Song}, {Beuzit}, {Lowrance}, \& {Bessell}}]{Chauvin2005}
{Chauvin}, G., {Lagrange}, A.-M., {Zuckerman}, B., {et~al.} 2005, \aap, 438,
  L29

\bibitem[{{Cumming} {et~al.}(2008){Cumming}, {Butler}, {Marcy}, {Vogt},
  {Wright}, \& {Fischer}}]{Cumming2008}
{Cumming}, A., {Butler}, R.~P., {Marcy}, G.~W., {et~al.} 2008, \pasp, 120, 531

\bibitem[{{Deming} {et~al.}(2009){Deming}, {Seager}, {Winn}, {Miller-Ricci},
  {Clampin}, {Lindler}, {Greene}, {Charbonneau}, {Laughlin}, {Ricker},
  {Latham}, \& {Ennico}}]{Deming2009}
{Deming}, D., {Seager}, S., {Winn}, J., {et~al.} 2009, \pasp, 121, 952

\bibitem[{{Dyudina} {et~al.}(2005){Dyudina}, {Sackett}, {Bayliss}, {Seager},
  {Porco}, {Throop}, \& {Dones}}]{Dyudina2005}
{Dyudina}, U.~A., {Sackett}, P.~D., {Bayliss}, D.~D.~R., {et~al.} 2005, \apj,
  618, 973

\bibitem[{{Fortney} {et~al.}(2008){Fortney}, {Marley}, {Saumon}, \&
  {Lodders}}]{Fortney2008}
{Fortney}, J.~J., {Marley}, M.~S., {Saumon}, D., \& {Lodders}, K. 2008, \apj,
  683, 1104

\bibitem[{{Fressin} {et~al.}(2013){Fressin}, {Torres}, {Charbonneau}, {Bryson},
  {Christiansen}, {Dressing}, {Jenkins}, {Walkowicz}, \&
  {Batalha}}]{Fressin2013}
{Fressin}, F., {Torres}, G., {Charbonneau}, D., {et~al.} 2013, \apj, 766, 81

\bibitem[{{Hansen} \& {Hovenier}(1974)}]{Hansen1974}
{Hansen}, J.~E., \& {Hovenier}, J.~W. 1974, in IAU Symposium, Vol.~65,
  Exploration of the Planetary System, ed. A.~{Woszczyk} \& C.~{Iwaniszewska},
  197--200

\bibitem[{{Howard} {et~al.}(2010){Howard}, {Marcy}, {Johnson}, {Fischer},
  {Wright}, {Isaacson}, {Valenti}, {Anderson}, {Lin}, \& {Ida}}]{Howard2010}
{Howard}, A.~W., {Marcy}, G.~W., {Johnson}, J.~A., {et~al.} 2010, Science, 330,
  653

\bibitem[{{Hu}(2014)}]{Hu2014}
{Hu}, R. 2014, ArXiv e-prints, arXiv:1412.7582

\bibitem[{{Irwin} {et~al.}(2002){Irwin}, {Calcutt}, {Weir}, {Taylor}, \&
  {Carlson}}]{Irwin2002}
{Irwin}, P.~G.~J., {Calcutt}, S.~B., {Weir}, A.~L., {Taylor}, F.~W., \&
  {Carlson}, R.~W. 2002, Advances in Space Research, 29, 285

\bibitem[{{Janson}(2010)}]{Janson2010}
{Janson}, M. 2010, \mnras, 408, 514

\bibitem[{{Kane}(2013)}]{Kane2013}
{Kane}, S.~R. 2013, \apj, 766, 10

\bibitem[{{Kane} \& {Gelino}(2010)}]{Kane2010}
{Kane}, S.~R., \& {Gelino}, D.~M. 2010, \apj, 724, 818

\bibitem[{{Kane} \& {Gelino}(2011)}]{Kane2011}
---. 2011, \apj, 729, 74

\bibitem[{{Karkoschka}(1994)}]{Karkoschka1994}
{Karkoschka}, E. 1994, \icarus, 111, 174

\bibitem[{{Karkoschka} \& {Tomasko}(2011)}]{Karkoschka2011}
{Karkoschka}, E., \& {Tomasko}, M.~G. 2011, \icarus, 211, 780

\bibitem[{{Kipping}(2013)}]{Kipping2013}
{Kipping}, D.~M. 2013, \mnras, 434, L51

\bibitem[{{Kuzuhara} {et~al.}(2013){Kuzuhara}, {Tamura}, {Kudo}, {Janson},
  {Kandori}, {Brandt}, {Thalmann}, {Spiegel}, {Biller}, {Carson}, {Hori},
  {Suzuki}, {Burrows}, {Henning}, {Turner}, {McElwain}, {Moro-Mart{\'{\i}}n},
  {Suenaga}, {Takahashi}, {Kwon}, {Lucas}, {Abe}, {Brandner}, {Egner}, {Feldt},
  {Fujiwara}, {Goto}, {Grady}, {Guyon}, {Hashimoto}, {Hayano}, {Hayashi},
  {Hayashi}, {Hodapp}, {Ishii}, {Iye}, {Knapp}, {Matsuo}, {Mayama}, {Miyama},
  {Morino}, {Nishikawa}, {Nishimura}, {Kotani}, {Kusakabe}, {Pyo}, {Serabyn},
  {Suto}, {Takami}, {Takato}, {Terada}, {Tomono}, {Watanabe}, {Wisniewski},
  {Yamada}, {Takami}, \& {Usuda}}]{Kuzuhara2013}
{Kuzuhara}, M., {Tamura}, M., {Kudo}, T., {et~al.} 2013, \apj, 774, 11

\bibitem[{{Lagrange} {et~al.}(2009){Lagrange}, {Gratadour}, {Chauvin}, {Fusco},
  {Ehrenreich}, {Mouillet}, {Rousset}, {Rouan}, {Allard}, {Gendron}, {Charton},
  {Mugnier}, {Rabou}, {Montri}, \& {Lacombe}}]{Lagrange2009}
{Lagrange}, A.-M., {Gratadour}, D., {Chauvin}, G., {et~al.} 2009, \aap, 493,
  L21

\bibitem[{{Macintosh} {et~al.}(2008){Macintosh}, {Graham}, {Palmer}, {Doyon},
  {Dunn}, {Gavel}, {Larkin}, {Oppenheimer}, {Saddlemyer}, {Sivaramakrishnan},
  {Wallace}, {Bauman}, {Erickson}, {Marois}, {Poyneer}, \&
  {Soummer}}]{Macintosh2008}
{Macintosh}, B.~A., {Graham}, J.~R., {Palmer}, D.~W., {et~al.} 2008, in Society
  of Photo-Optical Instrumentation Engineers (SPIE) Conference Series, Vol.
  7015, Society of Photo-Optical Instrumentation Engineers (SPIE) Conference
  Series, 18

\bibitem[{{Madhusudhan} \& {Burrows}(2012)}]{Madhu2012}
{Madhusudhan}, N., \& {Burrows}, A. 2012, \apj, 747, 25

\bibitem[{{Marley} {et~al.}(2014){Marley}, {Lupu}, {Lewis}, {Line}, {Morley},
  \& {Fortney}}]{Marley2014}
{Marley}, M., {Lupu}, R., {Lewis}, N., {et~al.} 2014, ArXiv e-prints,
  arXiv:1412.8440

\bibitem[{{Marley} {et~al.}(2007){Marley}, {Fortney}, {Hubickyj},
  {Bodenheimer}, \& {Lissauer}}]{Marley2007}
{Marley}, M.~S., {Fortney}, J.~J., {Hubickyj}, O., {Bodenheimer}, P., \&
  {Lissauer}, J.~J. 2007, \apj, 655, 541

\bibitem[{{Marley} {et~al.}(1999){Marley}, {Gelino}, {Stephens}, {Lunine}, \&
  {Freedman}}]{Marley1999}
{Marley}, M.~S., {Gelino}, C., {Stephens}, D., {Lunine}, J.~I., \& {Freedman},
  R. 1999, \apj, 513, 879

\bibitem[{{Marois} {et~al.}(2008){Marois}, {Macintosh}, {Barman}, {Zuckerman},
  {Song}, {Patience}, {Lafreni{\`e}re}, \& {Doyon}}]{Marois2008}
{Marois}, C., {Macintosh}, B., {Barman}, T., {et~al.} 2008, Science, 322, 1348

\bibitem[{{Marois} {et~al.}(2010){Marois}, {Zuckerman}, {Konopacky},
  {Macintosh}, \& {Barman}}]{Marois2010}
{Marois}, C., {Zuckerman}, B., {Konopacky}, Q.~M., {Macintosh}, B., \&
  {Barman}, T. 2010, \nat, 468, 1080

\bibitem[{{Neuh{\"a}user} {et~al.}(2005){Neuh{\"a}user}, {Guenther},
  {Wuchterl}, {Mugrauer}, {Bedalov}, \& {Hauschildt}}]{GQLupb}
{Neuh{\"a}user}, R., {Guenther}, E.~W., {Wuchterl}, G., {et~al.} 2005, \aap,
  435, L13

\bibitem[{{Nielsen} \& {Close}(2010)}]{Nielsen2010}
{Nielsen}, E.~L., \& {Close}, L.~M. 2010, \apj, 717, 878

\bibitem[{{Oppenheimer} \& {Hinkley}(2009)}]{Oppenheimer2009}
{Oppenheimer}, B.~R., \& {Hinkley}, S. 2009, \araa, 47, 253

\bibitem[{{Perryman} {et~al.}(2014){Perryman}, {Hartman}, {Bakos}, \&
  {Lindegren}}]{Perryman2014}
{Perryman}, M., {Hartman}, J., {Bakos}, G.~{\'A}., \& {Lindegren}, L. 2014,
  \apj, 797, 14

\bibitem[{{Satoh} {et~al.}(2000){Satoh}, {Itoh}, {Kawabata}, {Tenma}, \&
  {Akabane}}]{Satoh2000}
{Satoh}, T., {Itoh}, S., {Kawabata}, K., {Tenma}, T., \& {Akabane}, T. 2000,
  \pasj, 52, 363

\bibitem[{{Seager} {et~al.}(2000){Seager}, {Whitney}, \&
  {Sasselov}}]{Seager2000}
{Seager}, S., {Whitney}, B.~A., \& {Sasselov}, D.~D. 2000, \apj, 540, 504

\bibitem[{{Sharp} \& {Burrows}(2007)}]{SharpBurrows2007}
{Sharp}, C.~M., \& {Burrows}, A. 2007, \apjs, 168, 140

\bibitem[{{Skrutskie} {et~al.}(2010){Skrutskie}, {Jones}, {Hinz}, {Garnavich},
  {Wilson}, {Nelson}, {Solheid}, {Durney}, {Hoffmann}, {Vaitheeswaran},
  {McMahon}, {Leisenring}, \& {Wong}}]{Skrutskie2010}
{Skrutskie}, M.~F., {Jones}, T., {Hinz}, P., {et~al.} 2010, in Society of
  Photo-Optical Instrumentation Engineers (SPIE) Conference Series, Vol. 7735,
  Society of Photo-Optical Instrumentation Engineers (SPIE) Conference Series,
  3

\bibitem[{{Sozzetti} {et~al.}(2014){Sozzetti}, {Giacobbe}, {Lattanzi},
  {Micela}, {Morbidelli}, \& {Tinetti}}]{Sozzetti2014}
{Sozzetti}, A., {Giacobbe}, P., {Lattanzi}, M.~G., {et~al.} 2014, \mnras, 437,
  497

\bibitem[{{Spergel} {et~al.}(2015){Spergel}, {Gehrels}, {Baltay}, {Bennett},
  {Breckinridge}, {Donahue}, {Dressler}, {Gaudi}, {Greene}, {Guyon}, {Hirata},
  {Kalirai}, {Kasdin}, {Macintosh}, {Moos}, {Perlmutter}, {Postman},
  {Rauscher}, {Rhodes}, {Wang}, {Weinberg}, {Benford}, {Hudson}, {Jeong},
  {Mellier}, {Traub}, {Yamada}, {Capak}, {Colbert}, {Masters}, {Penny},
  {Savransky}, {Stern}, {Zimmerman}, {Barry}, {Bartusek}, {Carpenter}, {Cheng},
  {Content}, {Dekens}, {Demers}, {Grady}, {Jackson}, {Kuan}, {Kruk}, {Melton},
  {Nemati}, {Parvin}, {Poberezhskiy}, {Peddie}, {Ruffa}, {Wallace}, {Whipple},
  {Wollack}, \& {Zhao}}]{Spergel2015}
{Spergel}, D., {Gehrels}, N., {Baltay}, C., {et~al.} 2015, ArXiv e-prints,
  arXiv:1503.03757

\bibitem[{{Spiegel} \& {Burrows}(2012)}]{Spiegel2012}
{Spiegel}, D.~S., \& {Burrows}, A. 2012, \apj, 745, 174

\bibitem[{{Stam} {et~al.}(2004){Stam}, {Hovenier}, \& {Waters}}]{Stam2004}
{Stam}, D.~M., {Hovenier}, J.~W., \& {Waters}, L.~B.~F.~M. 2004, \aap, 428, 663

\bibitem[{{Sudarsky} {et~al.}(2003){Sudarsky}, {Burrows}, \&
  {Hubeny}}]{Sudarsky2003}
{Sudarsky}, D., {Burrows}, A., \& {Hubeny}, I. 2003, \apj, 588, 1121

\bibitem[{{Sudarsky} {et~al.}(2005){Sudarsky}, {Burrows}, {Hubeny}, \&
  {Li}}]{Sudarsky2005}
{Sudarsky}, D., {Burrows}, A., {Hubeny}, I., \& {Li}, A. 2005, \apj, 627, 520

\bibitem[{{Sudarsky} {et~al.}(2000){Sudarsky}, {Burrows}, \&
  {Pinto}}]{Sudarsky2000}
{Sudarsky}, D., {Burrows}, A., \& {Pinto}, P. 2000, \apj, 538, 885

\bibitem[{{Suzuki} {et~al.}(2010){Suzuki}, {Kudo}, {Hashimoto}, {Carson},
  {Egner}, {Goto}, {Hattori}, {Hayano}, {Hodapp}, {Ito}, {Iye}, {Jacobson},
  {Kandori}, {Kusakabe}, {Kuzuhara}, {Matsuo}, {Mcelwain}, {Morino}, {Oya},
  {Saito}, {Shelton}, {Stahlberger}, {Suto}, {Takami}, {Thalmann}, {Watanabe},
  {Yamada}, \& {Tamura}}]{Suzuki2010}
{Suzuki}, R., {Kudo}, T., {Hashimoto}, J., {et~al.} 2010, in Society of
  Photo-Optical Instrumentation Engineers (SPIE) Conference Series, Vol. 7735,
  Society of Photo-Optical Instrumentation Engineers (SPIE) Conference Series,
  30

\bibitem[{{Traub} \& {Oppenheimer}(2010)}]{Truab2010}
{Traub}, W.~A., \& {Oppenheimer}, B.~R. 2010, {Direct Imaging of Exoplanets},
  ed. S.~{Seager}, 111--156

\end{thebibliography}

\appendix
\section{}
\vspace{-.3cm}
\numberwithin{equation}{section}
The orbital phase angle ($\alpha$) is related to the true anomaly ($\theta$) and the Keplerian elements by\footnote{Note $\Omega$ is necessary when an absolute celestial frame is designated. However, one can assume $\Omega=90^\circ$ or simply ignore this parameter when the system itself is allowed to provide a natural orientation.}
\begin{align}\label{eqn:phase_eqn}
\cos\alpha &= \sin(\theta + \omega_p) \sin i \sin\Omega - \cos\Omega \cos(\theta + \omega_p),\nonumber\\
&= \sin(\theta + \omega_p) \sin i,
\end{align}
where $i$ is the orbital inclination, $\omega_p$ is the argument of periastron, and $\Omega$ is the longitude of the ascending node. The second equality in Equation~(\ref{eqn:phase_eqn}) comes from the assumption that $\Omega=90^\circ$. For the orbit orientation, we use the convention $\alpha=\theta=0^\circ$ when $\omega_p=i=90^\circ$. The range of observable phase angles depends only on the inclination of the orbit: $90^\circ - i < \alpha < 90^\circ + i$. For eccentric orbits, the light curve will vary even when $i=0^\circ$. This is due to changes in the orbital distance ($d$) along the orbit, which is given by
\beq\label{eqn:distance}
d = \frac{a (1 - e^2)}{1 + e \cos\theta}\,.
\eeq
The Keplerian elements are connected to time ($t$) through the mean anomaly ($M=2\pi(t - t_p)/P$, where $t_p$ is the time of periastron passage), which is related to the eccentric anomaly ($E$) by Kepler's equation:
\beq
M = E - e \sin E.
\eeq
$E$ can be expressed in terms of the true anomaly as
\beq
\sin E = \frac{\sin\theta \sqrt{1 - e^2}}{1 + e \cos \theta}.
\eeq  
We can then relate the planet's true anomaly to time:
\begin{align}\label{eqn:time_eqn}
\frac{t-t_p}{P} = \frac{1}{2\pi}\Bigg[&-\frac{e \sin\theta \sqrt{1 - e^2}}{1 + e\cos \theta} \nonumber  \\
&+\,2 \tan^{-1}\Bigg(\sqrt{\frac{(1 - e)}{(1 + e)}}\tan\frac{\theta}{2}\Bigg) \Bigg].
\end{align}
Thus, for an arbitrary orbit orientation, Equations~(\ref{eqn:phase_eqn}) and (\ref{eqn:time_eqn}) combine to yield the exact orbital phase at any time.

\end{document}